\documentclass[twoside,twocolumn,9pt]{article}
\usepackage{extsizes}
\usepackage[super,sort&compress,comma]{natbib} 
\usepackage[left=1.5cm, right=1.5cm, top=1.785cm, bottom=2.0cm]{geometry}
\usepackage{balance}
\usepackage{mathptmx}
\usepackage{sectsty}
\usepackage{graphicx} 
\usepackage{lastpage}
\usepackage[format=plain,justification=justified,singlelinecheck=false,font={stretch=1.125,small,sf},labelfont=bf,labelsep=space]{caption}
\usepackage{float}
\usepackage{fancyhdr}
\usepackage{fnpos}
\usepackage[english]{babel}
\addto{\captionsenglish}{%
  
}
\usepackage{array}
\usepackage{droidsans}
\usepackage{charter}
\usepackage[T1]{fontenc}
\usepackage[usenames,dvipsnames]{xcolor}
\usepackage{setspace}
\usepackage[compact]{titlesec}
\usepackage{hyperref}

\usepackage{amsmath}

\definecolor{cream}{RGB}{222,217,201}

\begin{document}

\pagestyle{fancy}
\thispagestyle{plain}
\fancypagestyle{plain}{
\renewcommand{\headrulewidth}{0pt}
}

\makeFNbottom
\makeatletter
\renewcommand\LARGE{\@setfontsize\LARGE{15pt}{17}}
\renewcommand\Large{\@setfontsize\Large{12pt}{14}}
\renewcommand\large{\@setfontsize\large{10pt}{12}}
\renewcommand\footnotesize{\@setfontsize\footnotesize{7pt}{10}}
\makeatother

\renewcommand{\thefootnote}{\fnsymbol{footnote}}
\renewcommand\footnoterule{\vspace*{1pt}%
\color{cream}\hrule width 3.5in height 0.4pt \color{black}\vspace*{5pt}} 
\setcounter{secnumdepth}{5}

\makeatletter 
\renewcommand\@biblabel[1]{#1}            
\renewcommand\@makefntext[1]%
{\noindent\makebox[0pt][r]{\@thefnmark\,}#1}
\makeatother 
\renewcommand{\figurename}{\small{Fig.}~}
\sectionfont{\sffamily\Large}
\subsectionfont{\normalsize}
\subsubsectionfont{\bf}
\setstretch{1.125} 
\setlength{\skip\footins}{0.8cm}
\setlength{\footnotesep}{0.25cm}
\setlength{\jot}{10pt}
\titlespacing*{\section}{0pt}{4pt}{4pt}
\titlespacing*{\subsection}{0pt}{15pt}{1pt}

\renewcommand{\headrulewidth}{0pt} 
\renewcommand{\footrulewidth}{0pt}
\setlength{\arrayrulewidth}{1pt}
\setlength{\columnsep}{6.5mm}
\setlength\bibsep{1pt}

\makeatletter 
\newlength{\figrulesep} 
\setlength{\figrulesep}{0.5\textfloatsep} 

\newcommand{\topfigrule}{\vspace*{-1pt}%
\noindent{\color{cream}\rule[-\figrulesep]{\columnwidth}{1.5pt}} }

\newcommand{\botfigrule}{\vspace*{-2pt}%
\noindent{\color{cream}\rule[\figrulesep]{\columnwidth}{1.5pt}} }

\newcommand{\dblfigrule}{\vspace*{-1pt}%
\noindent{\color{cream}\rule[-\figrulesep]{\textwidth}{1.5pt}} }

\makeatother

\twocolumn[
  \begin{@twocolumnfalse}
\begin{tabular}{m{4.5cm} p{13.5cm} }

& \noindent\LARGE{\textbf{An experimentally validated neural-network potential energy surface for H atoms on free-standing graphene in full dimensionality}} \\
\vspace{0.3cm} & \vspace{0.3cm} \\

 & \noindent\large{Sebastian Wille\textit{$^{a,c}$}, Hongyan Jiang\textit{$^{a}$}, Oliver B\"unermann\textit{$^{a,b,d}$}, Alec~M.~Wodtke\textit{$^{a,b,d}$}, J\"org Behler\textit{$^{c,d}$}, and Alexander Kandratsenka$^{\ast}$\textit{$^{a}$}} \\
\\
& \noindent\normalsize{We present a first principles-quality potential energy surface (PES) describing the inter-atomic forces for hydrogen atoms interacting with free-standing graphene. The PES is a high-dimensional neural network potential that has been parameterized to 75\,945 data points computed with density-functional theory employing the PBE-D2 functional. Improving over a previously published PES [Jiang \textit{et al., Science}, 2019, \textbf{364}, 379], this neural network exhibits a realistic physisorption well and achieves a 10-fold reduction in the RMS fitting error, which is 0.6\,meV/atom. 
We used this PES to calculate about 1.5 million classical trajectories with carefully selected initial conditions to allow for direct comparison to results of H- and D-atom scattering experiments performed at incidence translational energy of 1.9\,eV and a surface temperature of 300\,K. The theoretically predicted scattering angular and energy loss distributions are in good agreement with experiment, despite the fact that the experiments employed graphene grown on Pt(111). The remaining discrepancies between experiment and theory are likely due to the influence of the Pt substrate only present in the experiment. } \\

\end{tabular}

 \end{@twocolumnfalse} \vspace{0.6cm}

  ]

\renewcommand*\rmdefault{bch}\normalfont\upshape
\rmfamily
\section*{}
\vspace{-1cm}


\footnotetext{\textit{$^{a}$~Department of Dynamics at Surfaces, Max-Planck-Institute for Biophysical Chemistry, Am Fa{\ss}berg 11, 37077 G{\"o}ttingen, Germany. E-mail: akandra@mpibpc.mpg.de}}
\footnotetext{\textit{$^{b}$~Institute for Physical Chemistry, Georg-August University of G\"ottingen, Tammann-stra\ss e 6, 37077 G\"ottingen, Germany}}
\footnotetext{\textit{$^{c}$~Institute for Physical Chemistry, Theoretische Chemie, Georg-August University of G\"ottingen, Tammannstra\ss e 6, 37077 G\"ottingen, Germany.}}
\footnotetext{\textit{$^{d}$~International Center for Advanced Studies of Energy Conversion, Georg-August University of G\"ottingen, Tammannstra\ss e 6, 37077 G\"ottingen, Germany.}}




\section{Introduction}							%
H-atom chemisorption to graphene is relevant to hydrogen storage \cite{Schlapbach2001}, the catalytic formation of molecular hydrogen in the interstellar medium \cite{Hornekaer2003} and {---} because hydrogenation of graphene can induce a band-gap {---} two-dimensional semiconductor materials \cite{Balog2010}. Recently, a full-dimensional PES was reported using first principles energies obtained from Embedded Mean-field Theory (EMFT) \cite{Fornace2015, Ding2017a, Ding2017b} to parameterize a second generation Reactive Empirical Bond Order (REBO) function\cite{Brenner2002}. Using classical and semi-classical dynamics calculations, qualitative agreement was obtained with H-atom scattering experiments carried out at incidence translational energies $E_i$ of  1.9 and 1\,eV. Furthermore, the trajectories provided an atomic scale movie at the femtosecond time scale showing the formation of a covalent chemical bond\cite{Jiang2019}. 
The sticking probability could also be calculated using the REBO-EMFT PES and compared well with experiment at $E_i = 1$\,eV. This suggests that the REBO-EMFT PES is the best available representation of interatomic forces in the H/graphene system. 

Despite the progress made in that work, two problems remained. First, the EMFT data was derived from a model of free-standing graphene, while the experiment was carried out on graphene that had been grown on Pt(111) \cite{Jiang2019}. To account for the influence of Pt in the simulations, a Lennard-Jones (LJ) interaction model with the Pt substrate was included for each atom in the graphene layer. This improved agreement with experiment, suggesting the influence of the substrate may be important. Unfortunately, it is unclear how to reparameterize the analytical REBO-EMFT PES from first-principles energies that include the Pt substrate. Therefore, the role of the substrate remains uncertain. 
The second problem concerns the fitting error (7\,meV/atom) as the REBO function is not flexible enough to closely reproduce electronic structure data\cite{Jiang2019}. 
This complicates the evaluation of the quality of different electronic structure methods, since the fitting error can easily be larger than the energy differences between the methods being compared.  Clearly, a full-dimensional first-principles PES where fitting errors are small and where the role of the Pt substrate is included would be a significantly better approach to this problem. For both of these problems a solution is offered by atomistic potentials employing machine learning (ML) methods.

In recent years, ML potentials have become a promising new approach to construct PESs of first-principles quality~\cite{P4885,P5793}. They have a uniquely flexible functional form that allows the accurate reproduction of reference data sets obtained in electronic structure calculations, without sacrificing the efficiency needed when they are repetitively evaluated in large-scale molecular dynamics (MD) simulations. ML potentials have been developed for many systems. These include free-standing~\cite{P5187} and multi-layer graphene~\cite{P5850} as well as graphite~\cite{P5850,P2594,P3007} and amorphous carbon~\cite{P4958}, which are closely related to this work. A frequently used type of machine learning potential suitable for large condensed systems is the high-dimensional neural network potential (HDNN-PES) method proposed by Behler and Parrinello in 2007~\cite{PRL2007}.

In this paper, we present the first HDNN-PES for H atoms interacting with free standing graphene, which we validate against data obtained from H and D scattering experiments using graphene grown on Pt(111), experiments that are similar to those recently reported elsewhere.\cite{jiang2020nuclear} Compared to the REBO-EMFT PES\cite{Jiang2019}, we achieve substantially reduced fitting errors without sacrificing computational performance. Using molecular dynamics, we show that experimentally obtained H/D-atom energy loss and angular distributions are faithfully reproduced. We demonstrate the improvement represented by the HDNN-PES by comparing the new results to MD simulations done with the previously reported REBO-EMFT PES\cite{Jiang2019}. The remaining deviations between experiment and theory likely reflect the absence of the Pt substrate in our simulations; however, the influence of the substrate on the scattering distributions appears to be relatively small.


\section{Experimental Methods}\label{sec:expmethods}					%

The experimental apparatus has been described in detail in Ref. \citenum{Buenermann2018}. H/D-atoms are generated by photodissociating a supersonic molecular beam of hydrogen/deuterium iodide with a KrF excimer laser producing atoms with incidence energy of 1.92\,eV. A small fraction of these atoms passes through two differential pumping stages, enter the ultra-high vacuum chamber and collide with the graphene sample grown \textit{in situ} on a Pt(111) substrate. The sample is held on a six-axis manipulator, allowing variation of the incidence angle $\theta_i$. Recoiling atoms are excited to a long lived Rydberg state ($n = 34$) by two laser pulses at 121.5\,nm and 365\,nm via a two-step excitation. These neutral atoms travel 25\,cm in a field-free region and pass a detector aperture before they are field-ionized and detected by a multi-channel plate detector. The arrival time is recorded by a multi-channel scalar. The rotatable detector allows data to be recorded at various scattering angles $\theta_s$. The graphene sample is epitaxially grown on a clean Pt(111) substrate by dosing ethylene (partial pressure $3\times10^{-8}$\,mbar) at $700^\circ$\,C for 15 minutes.

\section{Computational Methods}\label{sec:computationaldetails}
\subsection{HDNN-PES} 
High-dimensional neural network potentials (HDNN-PESs)~\cite{PRL2007} have been the first type of ML potential enabling the simulations of large condensed systems. In this approach, the total potential energy $E_\text{tot}$ of the system is constructed as a sum of atomic energy contributions,
\begin{equation}
E_\text{tot} = \sum^{N_\text{atoms}}_{\mu = 1} E_\mu,
\label{eq:poster}
\end{equation}
depending on the local chemical environment defined by a cutoff radius $R_\text{c}$, typically in the range between 6 and 10\,\AA. The positions of all neighboring atoms inside the cutoff spheres are described by sets of atom-centered many-body symmetry functions~\cite{P2882}. The resulting vector of symmetry function values for each atom represents a structural fingerprint that is used as input for an atomic neural network yielding atomic energy contribution $E_{\mu}$ into the total energy (\ref{eq:poster}). The functional forms of the symmetry functions ensure the necessary invariance of PES with respect to  translations and rotations of the system as well as permutations of like atoms. The atomic neural networks are feed-forward neural networks and contain a large number of weight parameters, which serve as fitting parameters for the HDNN-PES. Each element in the system is modeled by a separate atomic neural network with a specific architecture and values of the weight parameters calculated once for each atom of the respective element in the system. The values of these parameters are determined in an iterative optimization process by minimizing the errors of the energies and forces for a reference data set of representative structures obtained from electronic structure calculations, typically density-functional theory. Additional structures that may be required in the reference set in regions of the PES that are not well sampled can be suggested by an automatic procedure employing a committee of HDNN-PESs and a comparison of predicted energies and forces~\cite{P3114} leading to a self-consistent and unbiased generation of the data set.
Once a set of weight parameters accurately reproducing the reference data  has been found, the PES undergoes a series of careful validation steps\cite{P4444}. Then, the HDNN-PES is ready for applications. 
For all the details about the HDNN-PES method, the determination of the weight parameters and its validation procedures, the interested reader is referred to several recent reviews~\cite{P5128,P4444,P4106}

\subsection{Density Functional Theory Calculations}

\begin{figure}[t!]
\includegraphics[width=1.0\columnwidth]{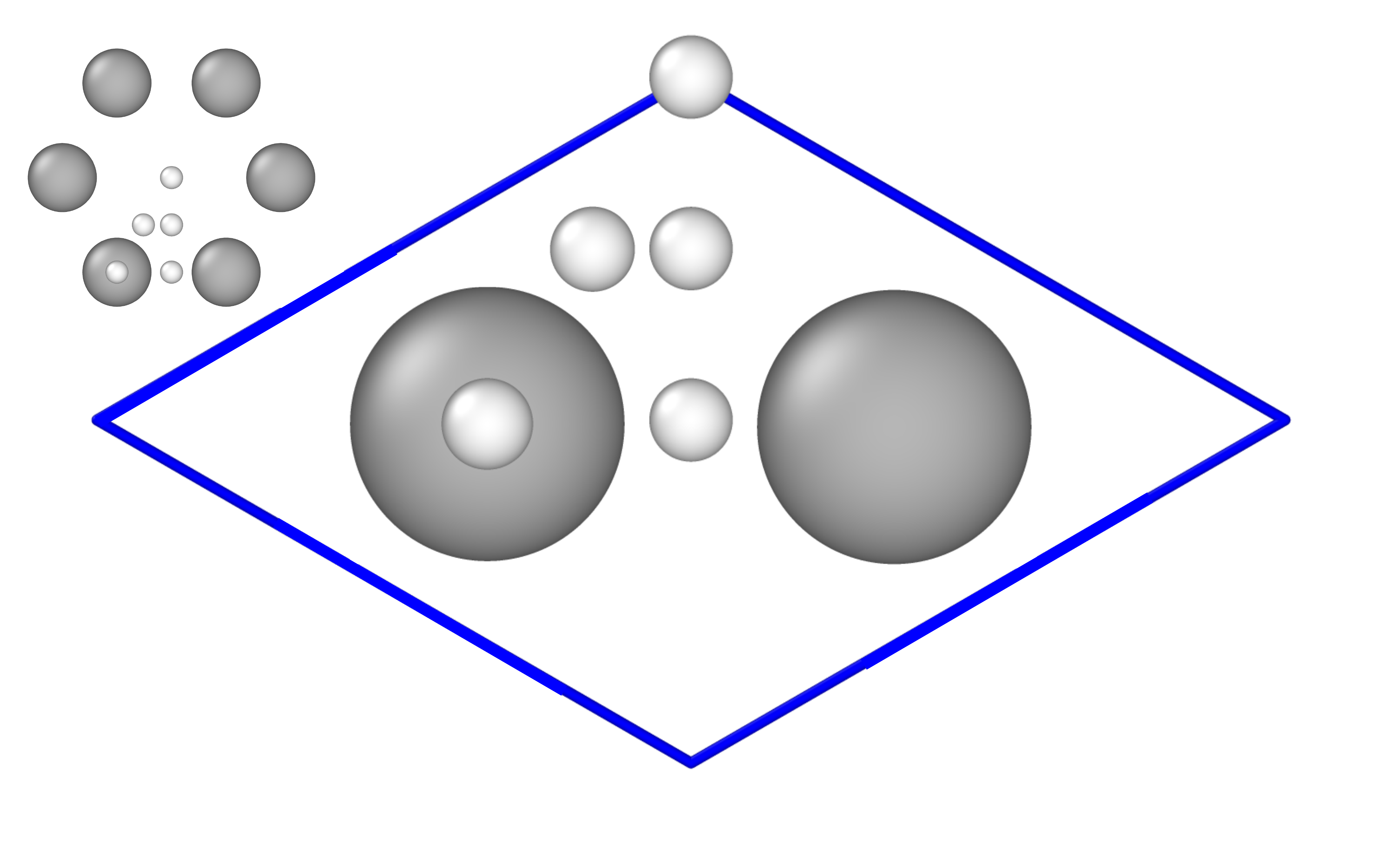}
    \caption{Primitive cell containing two C-atoms used to create the $3\times 4$ graphene slab. Important high-symmetry sites are indicated by small white balls. This figure and the ones showing surface structures of graphene are created using OVITO version 2.9.0\cite{ovito}}\label{Fig:unitcell}
\end{figure}

The Vienna \textit{ab initio} simulation package (VASP) version 5.3.5~\cite{P0156,Kresse1995,Kresse1996,Kresse1999} has been employed for the reference electronic structure calculations to generate the training set for the HDNN-PES. Density functional theory (DFT) at the generalized gradient approximation (GGA) level of theory using the Perdew-Burke-Ernzerhof (PBE)~\cite{Perdew1996a} exchange-correlation functional with a plane-wave basis has been used in combination with Grimme D2 van der Waals (vdW) corrections~\cite{Grimme2006SemiempiricalGD}. We made use of the Projector Augmented Wave (PAW)~\cite{Kresse1999,PhysRevB.50.17953} approach to model the core and valence electron interactions. The kinetic energy cutoff has been set to 400\,eV. The Monkhorst-Pack scheme~\cite{Monkhorst1976} with a $8\times 8\times 1$ $\Gamma$-centered $k$-point mesh for the $3\times 4$ surface cell has been used to sample the surface Brillouin zone. With two atoms per primitive unit cell, the slab consists of 24 carbon atoms in total and is 8.55\,\AA{} $\times$ 7.40\,\AA{} in size (see Fig.~\ref{Fig:unitcell}). 3D periodic boundary conditions have been applied with 13\,\AA vacuum perpendicular to the graphene sheet to ensure that the periodic images of the surfaces are non-interacting and that hydrogen atoms can be included at a maximum separation of 6\,\AA{} from the surface. 
We included spin polarization in the electronic structure calculations, and partial occupations have been treated by applying the tetrahedron method with Bl\"ochl corrections~\cite{PhysRevB.49.16223,PhysRevB.50.17953} using the default value of 0.2\,eV as the smearing parameter. The threshold for the change in energy between iteration steps when relaxing the electronic degrees of freedom has been $10^{-5}$\,eV.

\subsection{Generating the Reference Structures}

The iterative procedure described in detail elsewhere~\cite{Behler2012Cu} was used to generate the reference data. Briefly, step by step new DFT energies and forces are added for geometries where the HDNN-PES fit does not show the desired accuracy or covers the full configurational space. These geometries are identified by comparing the results of several generated HDNN-PESs with differing network structures. The reference data set is then extended with the additional data until convergence is reached.  

The initial data set consists of energies and forces obtained from DFT calculations for about $6\times10^4$ reference configurations, which were picked up from: (i) \textit{ab initio} molecular dynamics (AIMD) trajectories simulating  H-atom scattering from graphene at incidence energy of  1.9\,eV and incidence angles of 34$^\circ$ and 52$^\circ$ at surface temperatures of 300\,K and 600\,K;
(ii) geometries close to the minimum energy path to adsorption, where the H-atom was put at the lateral position of the
C-atom and the $z$-coordinates were varied over a range of $-0.8\,\text{\AA} \leq \text{H}_z\leq5.8$\,\AA{} and $-0.8\,\text{\AA}\leq\text{C}_z\leq 1.0$\,\AA{}, respectively, with  0.025\,\AA{} step and without structures with $r_\text{CH}<0.6$\,\AA{}, whereas the remaining C-atoms were kept at their equilibrium positions; (iii)  graphene geometries chosen randomly from an AIMD trajectory thermalized at 300\,K with H-atom over the surface. The position of the H-atom is chosen randomly over the whole simulation cell where the $z$-coordinate ranges from 1 to 6\,\AA. The configuration with a C--H distance of 6\,\AA{} and a fully relaxed graphene surface was used as the asymptotic energy reference. This structure is our energy zero point.


The HDNN-PESs fitted to the initial reference data set were then improved on the set of about $1.5\times10^4$ configurations obtained from  MD simulations of H-atom scattering from a graphene sheet at incidence energy of 1.9\,eV in the wide range of incidence angles (from 0$^\circ$ to 90$^\circ$ in 10$^\circ$ step) as well as at incidence energy of 6\,eV and normal incidence angle with surface temperatures of 0\,K and 600\,K starting over high-symmetry sites shown in Fig.~\ref{Fig:unitcell}. Moreover, we also trained the HDNN-PESs on the configurations taken from equilibrium MD simulations of the graphene surface in the wide range of temperatures from 0\,K to 2000\,K. The high-temperature configurations are useful, since  the surface can be heated locally in the neighborhood of the collision site.

In total, the final HDNN-PES was trained on the reference data set of 75\,945 configurations. 





\subsection{Construction of the Neural Network Potential}

The HDNN-PES has been constructed using the RuNNer\cite{P5128,P4444,IPClink} code. The atomic neural network's architecture consists of two hidden layers with 15 neurons per layer providing the energies both for hydrogen and carbon atoms. The parameters of symmetry functions\cite{P2882} are listed in Table~1 of SI. The symmetry function values have been rescaled to the range from 0 to 1.  Randomly selected 90\% of the reference data were used to train the NN, whereas the remaining 10\% were used as an independent test set to  validate the fit and to check for overfitting. The NN weight parameters were determined from the DFT energies and forces employing the adaptive global extended Kalman filter~\cite{P1308}. The initial values of the weight parameters have been chosen randomly in the interval from $-1$ to 1. For the weights, a preconditioning scheme was applied to reduce the initial root-mean-square error (RMSE)~\cite{P4444}. The training data in each of the 200 iterations (epochs) of the fit were presented in a random order to reduce the probability of getting trapped in local minima.

    

\subsection{Molecular Dynamics Simulations Details}

MD simulations of H-atom scattering from graphene were performed using MDT2 code\cite{MDT2GIT} developed to study atomic scattering from various surfaces\cite{Janke2015, Buenermann2015, Jiang2019}. The RuNNer subroutines implementing the HDNN-PES providing the energies and forces  were integrated into the MDT2 code. All the results shown in this paper have been obtained from MD simulations carried out using the RuNNer-MDT2 interface.

MD simulations of the H/D scattering from graphene have been carried out in the 
NVE ensemble using the standard velocity Verlet algorithm~\cite{allen1989computer,Swope1982} with a time step of 0.1\,fs. The trajectories were started with an H-atom randomly put at height of 3.5\,\AA{} over the surface and were terminated either when the scattered atom distance from the surface became larger than 3.6\,\AA{} or when the trajectory duration exceeded 200\,fs. The initial geometry for the graphene layer was randomly selected from 1000 configurations obtained after the equilibration of the surface at 300\,K with Andersen thermostat~\cite{dfrenkel96:mc,Andersen1980}. Those configurations were extracted from a single 100\,ps-long trajectory with a period of 100\,fs.
In the experiment, graphene is not a single crystal but a composition of two equally abundant orientational domains. The two domains have a rotational distribution with a Gaussian width of 5$^\circ$ and they are rotated by 27$^\circ$ with respect to each other\cite{Jiang2019}. This results in a H-atom velocity vector that is oriented symmetrically with respect to the two domains. To achieve scattering conditions comparable to experiment, the simulations have been carried out by averaging over two domains with incidence azimuth $\phi_i = \pm13.5^{\circ}$, where zero for azimuth angle is aligned with a C{=}C bond in graphene.
In total, $\sim$1.5 million trajectories have been carried out for different incidence angles and isotopes (hydrogen and deuterium). The exact numbers of trajectories for the different conditions can be found in Table~2 of the SI.

\section{Results}\label{sec:results}					%

\begin{figure}[t!]

\includegraphics[height=1.0\columnwidth,angle=90]{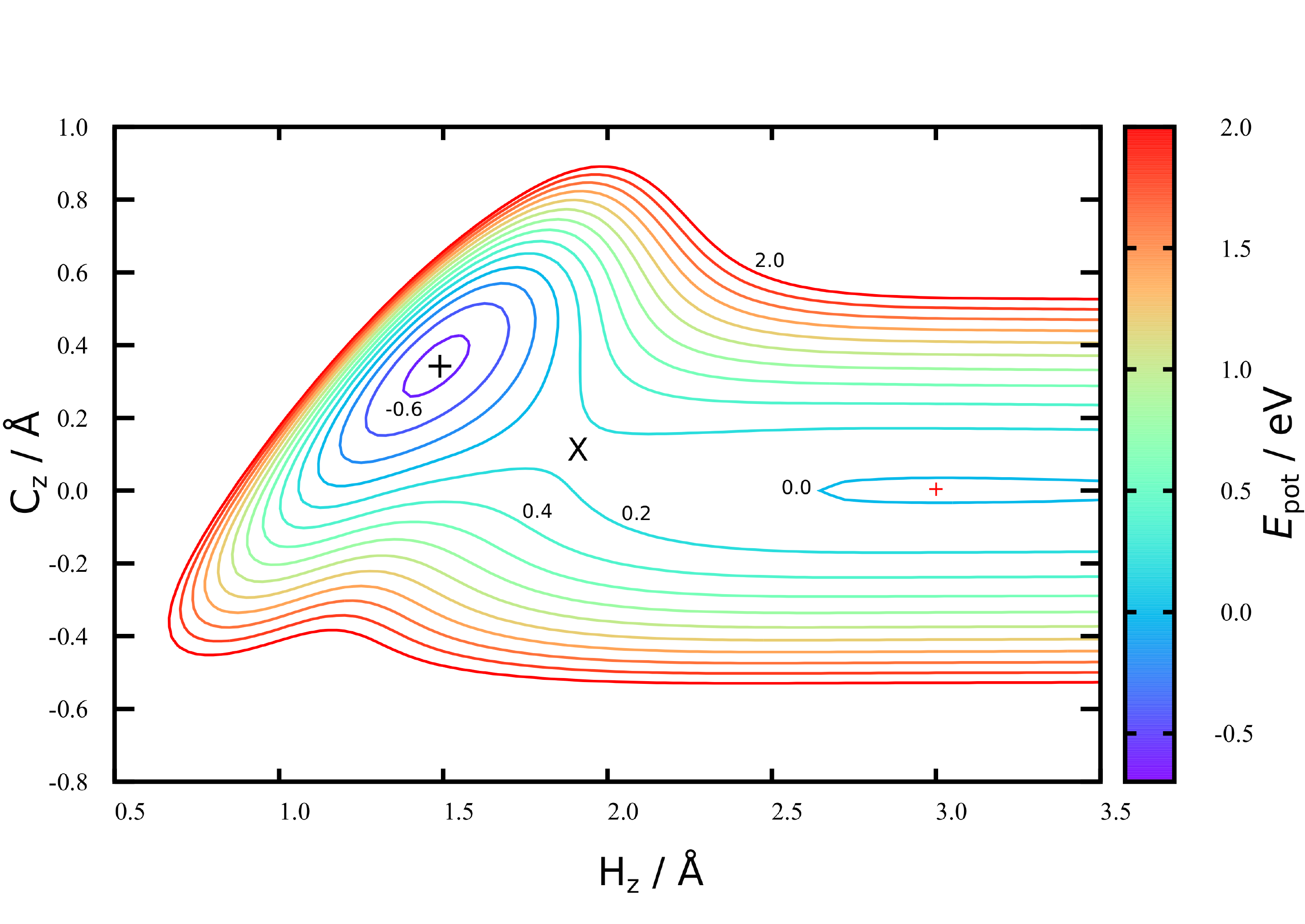}
    \caption{A cut through the HDNN-PES in the vicinity of the minimum energy path to chemisorption. The H atom is constrained to lie directly above a C-atom. H$_z$ and C$_z$ indicate the distance of H and C, respectively, from the plane of the graphene sheet. The physisorption (\textcolor{red}{+}) and chemisorption (+) wells have depths of 9 and 657\,meV, respectively. The barrier to chemisorption (\textbf{$\times$}) has a height of 172\,meV.}\label{Fig:PES_z}
\end{figure}

Fig.~\ref{Fig:PES_z} shows a two dimensional cut through the converged HDNN-PES developed in this work reflecting structures near the minimum energy path to chemisorption, where the H-atom approaches directly above a C-atom. A physisorption well can be seen at large H$_z$ and a deeper chemisorption well at small H$_z$ with C$_z\approx0.4$\,\AA. The minimum energy path to chemisorption involves both degrees of freedom, demonstrating that the C-atom is partially re-hybridized from sp$^2$ to sp$^3$ at the transition state. 

The depth of the physisorption well of the HDNN-PES has a depth of 9\,meV. This compares well with the DFT energy calculated at this configuration 22\,meV. The global physisorption minimum has the H-atom centered over the 6-ring. Here, the HDNN-PES gives the well depth of 11\,meV at an H$_z=2.7$\,\AA{}. This compares reasonably well with the experimentally determined physisorption well depth (40\,meV) \cite{ghio1980vibrational} and a correlated, counterpoise corrected wave function calculation of the hydrogen-coronene system, which found the minimum at H$_z=2.93$\,\AA{}
\cite{Bonfanti2007}. The previous REBO-EMFT PES had no physisorption well. 

The chemisorption well depth of the HDNN-PES (657\,meV) also compares well with DFT (676\,meV) but is deeper than that of the REBO-EMFT PES (610\,meV). Furthermore, the DFT barrier (160\,meV) is reproduced well by the HDNN-PES (172\,meV) but is lower than that of the REBO-EMFT PES (260\,meV).

\begin{figure}[t!]
\includegraphics[width=1.0\columnwidth]{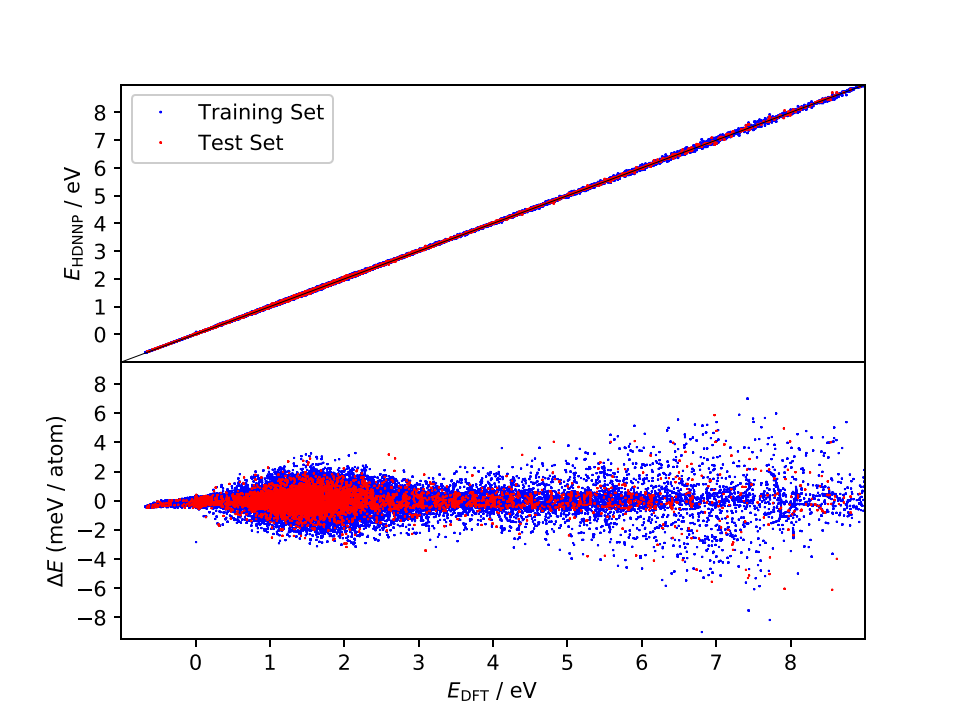}
    \caption{Fitting error of $E_\text{HDNN-PES}$ to $E_\text{DFT}$. The upper panel shows the comparison of the two energies and lower panel shows the signed error. DFT energy scale has its zero at configuration corresponding to a relaxed graphene sheet at $T = 0$\,K with an H atom 6\,\AA{} away from the plane of the graphene.}
\label{Fig:fiterr}
\end{figure}

The improved quality of the HDNN-PES in comparison to the REBO-EMFT PES is due both to the use of a dispersion corrected functional as well as to reduced fitting error. The RMSE fitting error of the REBO function to the DFT training data was reported to be $\approx7$\,meV/atom\cite{Jiang2019}; furthermore, the REBO function cannot represent a physisorption well. The flexibility of the neural network{---}the RMSE for the HDNN-PES is $\approx0.6$\,meV/atom for energies in training and test set and $\approx90$\,meV/\AA{} for forces in training and test set, respectively{---}easily leads to a physically realistic physisorption well and a more accurate representation of the DFT energies and forces. Fig.~\ref{Fig:fiterr} shows the fitting error to the DFT energies graphically. While the errors are not randomly distributed, there is no reason to suspect systematic problems with the PES over the energy range of 10\,eV.

\begin{figure}[t!]
\includegraphics[width=1.0\columnwidth]{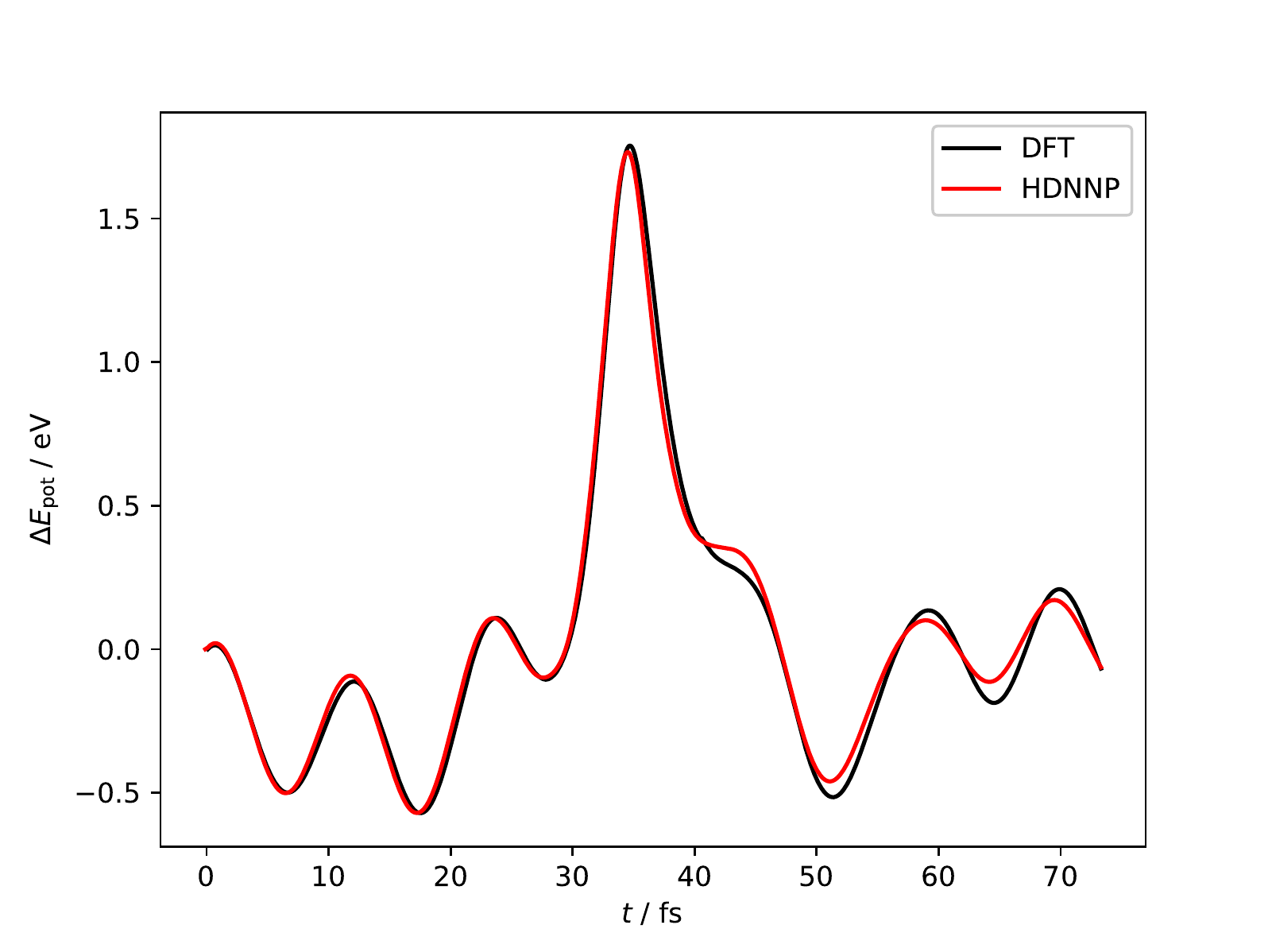}
    \caption{Potential energies from AIMD and HDNN-PES trajectories with $E_i=1.9$\,eV, $\theta_i =34^{\circ}$ and $\phi_i = 0^{\circ}$. The two trajectories were launched with identical initial conditions and both traverse the chemisorption well before returning to the gas phase after a single bounce.  The distance of closest  approach is below $r_\text{CH} =1.4$\,\AA. Movies of the two trajectories can be found in the SI. 
    }
    \label{FIG:AIMDNRG}
\end{figure}

\begin{figure}[t!]
\includegraphics[width=1.0\columnwidth]{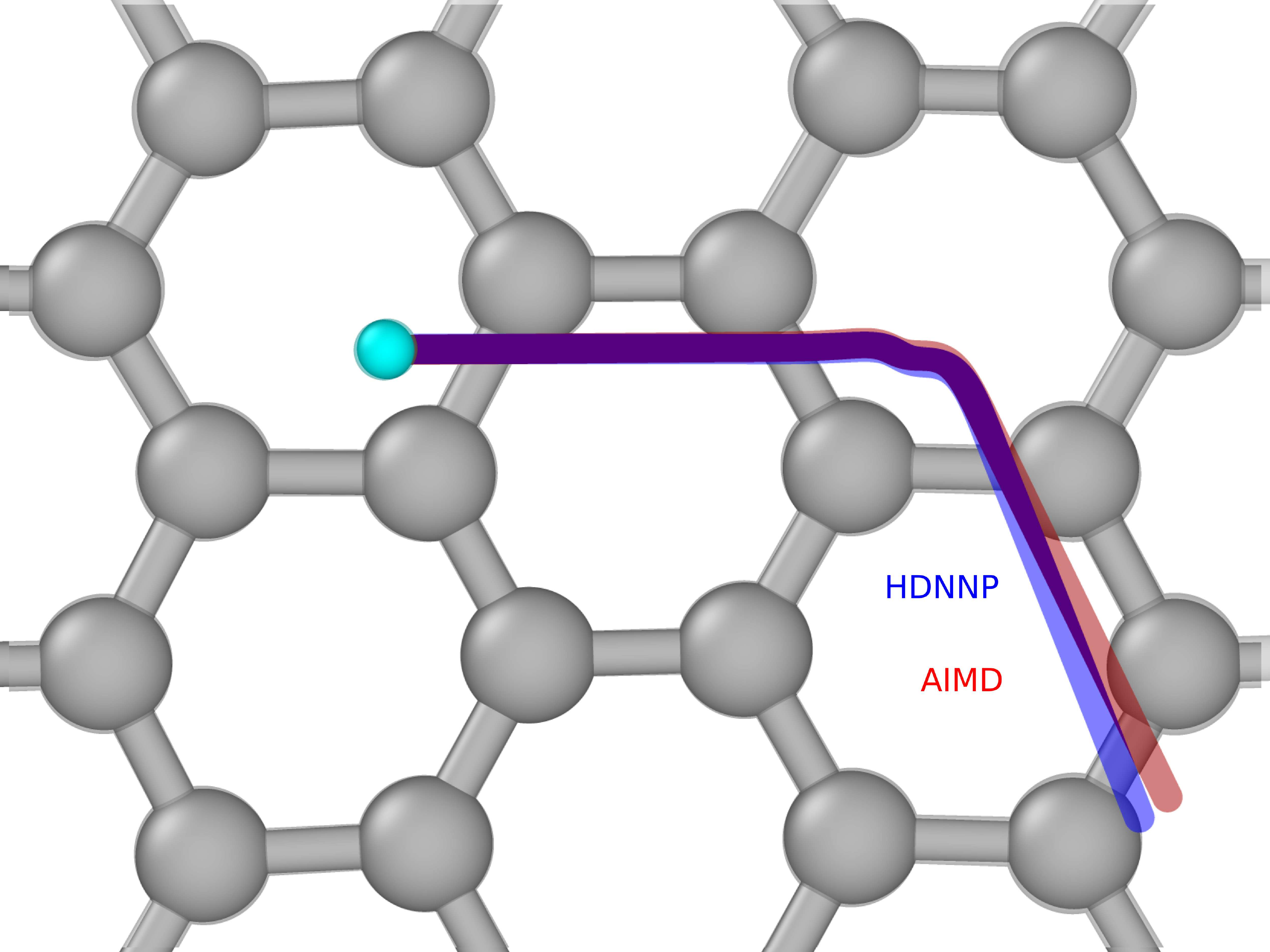}
    \caption{The same two trajectories as in Fig.~\ref{FIG:AIMDNRG}{---}AIMD (red) and HDNN-PES (blue). The H-atom's initial position is shown as a cyan colored ball. The divergence between the two trajectories is due to residual error in the NN fit to the DFT data. A "side view" of the trajectories can be found in the SI.}
    \label{FIG:AIMDTRAJ}
\end{figure}

Figs.~\ref{FIG:AIMDNRG} and \ref{FIG:AIMDTRAJ} show perhaps in the most impressive way the quality of the NN fitting. Here, two classical trajectories are represented, one performed with the HDNN-PES and one with AIMD. The trajectories correspond to the same initial conditions and are typical of those that will be compared to experiment below. Fig.~\ref{FIG:AIMDNRG} shows the potential energy change along the trajectory while Fig.~\ref{FIG:AIMDTRAJ} shows the H atom motion in the trajectory in a perspective drawing. The trajectories obtained with these two approaches are nearly identical{---}note that if the fitting were perfect they would be identical. We now turn to the question: How well is the experiment reproduced by classical MD on the new HDNN-PES.

\begin{figure}[t!]
\includegraphics[width=1.0\columnwidth]{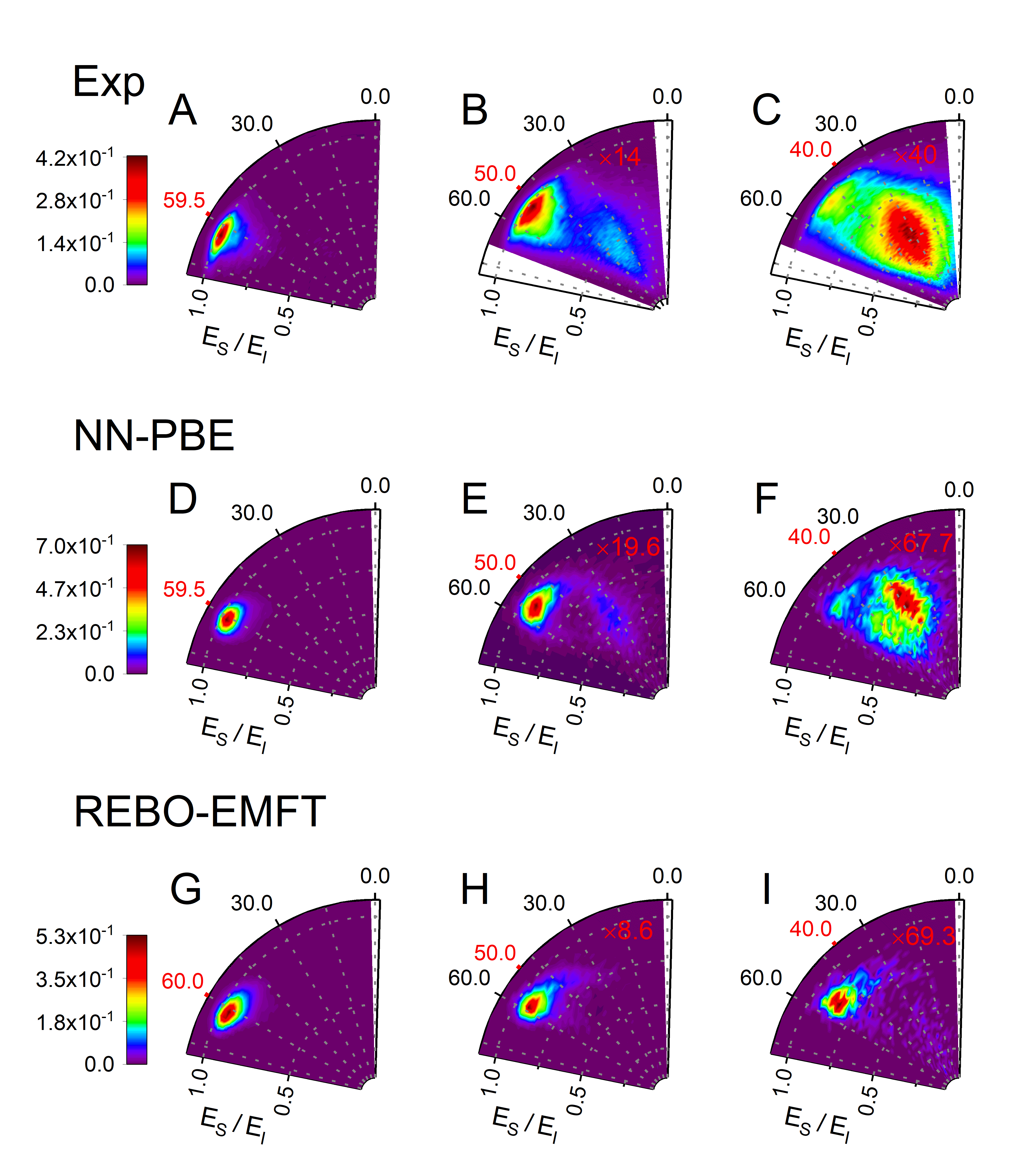}
    \caption{Comparing Theory with Experiment for H-scattering from graphene at incidence kinetic energy $E_i=1.9$\,eV. The energy loss is the fraction of $E_i$ and the kinetic energy of the H-atom after its collision with the surface $E_s$. Experimental distribution are shown in panels (A--C) along with theoretical distribution found from MD simulations using the HDNN-PES (D--F) and  the REBO-EMFT PES from Ref.~\citenum{Jiang2019} (G--H). The red labeled ticks indicate both the incidence and specular scattering angles. The integrated signals of panel A, D and G are normalized to 1. The number of trajectories used for the plots are shown in Table~2 in the SI.}
    \label{Fig:Hangendist}
\end{figure}

\begin{figure}[t!]
\includegraphics[width=1.0\columnwidth]{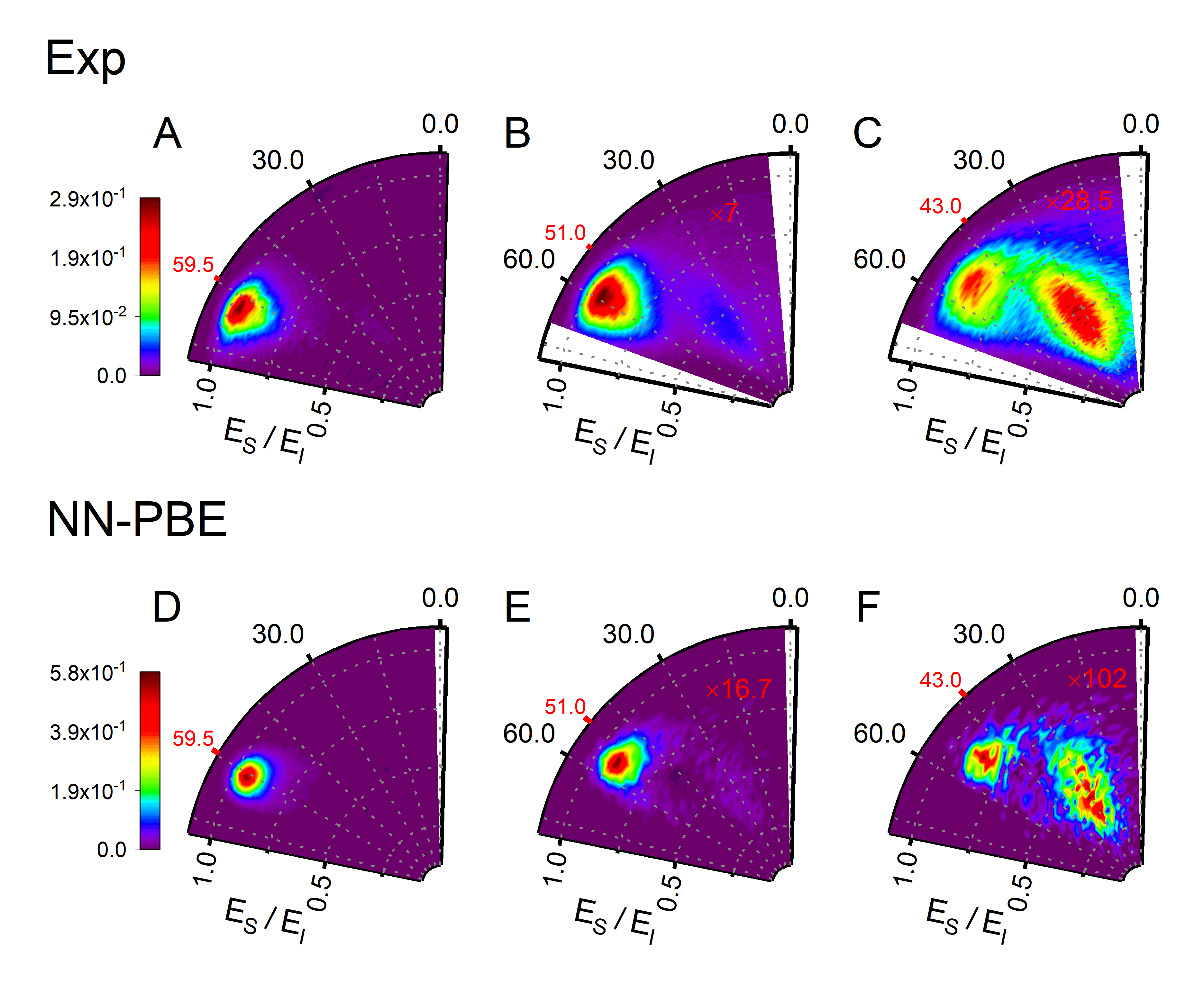}
    \caption{Comparing Theory with Experiment for D-scattering from graphene at $E_i=1.9$\,eV. Experimental distribution are shown in panels (A--C) along with theoretical distribution found from MD simulations using the HDNN-PES (D--F). The number of trajectories used for the plots are shown in Table~2 in the SI.\label{Fig:Dangendist} }
\end{figure}

Figs.~\ref{Fig:Hangendist} and \ref{Fig:Dangendist} show comparisons between experiment and theory for H and D scattering from graphene, respectively. In both figures, panels (A--C) show experimentally derived angle-resolved energy loss distributions represented as heat maps for three values of the incidence polar angle $\theta_i$ indicated with red numbers on the polar axes. The energy loss is the fraction of the incidence kinetic energy of the projectile $E_i$ and the kinetic energy after its collision with the surface $E_s$. Panels (D--F) show theoretically predicted distributions derived with the HDNN-PES of this work. In Fig.~\ref{Fig:Hangendist} we also show the distributions obtained when using the REBO-EMFT PES from Ref. \citenum{Jiang2019}{---}see panels (G--I).

The total observed scattering flux decreases rapidly as the incidence angle approaches the normal{---}this occurs for two reasons. First, the normal component of H/D kinetic energy is more effective in promoting passage over the barrier to chemisorption\cite{Jiang2019}. Thus, smaller incidence angles produce more sticking. Secondly, the experiment can only observe scattered atoms within a plane defined by the direction of the atomic beam and the normal to the surface. Changing the incidence angle affects the fraction of atoms scattered within that plane. The drop in scattering flux caused by the reduction of the incidence angle is indicated quantitatively by the multiplying factors on the panels. Clearly, the HDNN-PES predictions are in better agreement with experiment than those of the REBO-EMFT PES{---}see Fig.~\ref{Fig:Hangendist}.

Both H and D scattering from graphene exhibit two distinct energy loss channels: a quasi-elastic and a high energy loss channel. The quasi-elastic channel comes from trajectories that fail to cross the barrier to chemisorption, whereas the high energy loss channel arises from trajectories that passed through the chemisorption well forming a transient C--H bond and subsequently returned to the gas phase \cite{Jiang2019}. The relative intensities of these two channels are also sensitive to incidence angle. The experiment shows that at large incidence angles{---}see Figs.~\ref{Fig:Hangendist} and \ref{Fig:Dangendist} panels (A){---}only quasi-elastic scattering is seen. At small incidence angles{---}panels (C){---}transient chemical bond formation dominates and at intermediate incidence angles{---}panels (B){---}both channels contribute to the scattering signal. The angle-resolved energy loss distributions obtained with the HDNN-PES{---}Fig.~\ref{Fig:Hangendist} panels (D--F){---}capture these experimental observations qualitatively better than those obtained with the REBO-EMFT PES{---}Fig.~\ref{Fig:Hangendist} panels (G--I).  

The influence of isotopic substitution on the energy loss spectra can serve as an additional test to validate the accuracy of the HDNN-PES. Comparing the upper panels of Figs.~\ref{Fig:Hangendist} and \ref{Fig:Dangendist} shows that the experimentally observed branching into the high energy{--}loss channel is somewhat smaller for D than for H under the same incidence conditions. Classical trajectories carried out on the HDNN-PES describe this isotope effect well. Even subtle difference in the angle-resolved energy loss distributions seen in experiment are captured in the trajectory calculations. Compare for example, panels C (experiment) and F (MD with HDNN-PES) of Figs.~\ref{Fig:Hangendist} and \ref{Fig:Dangendist}. 

\begin{figure}[h!]
\includegraphics[width=1.0\columnwidth]{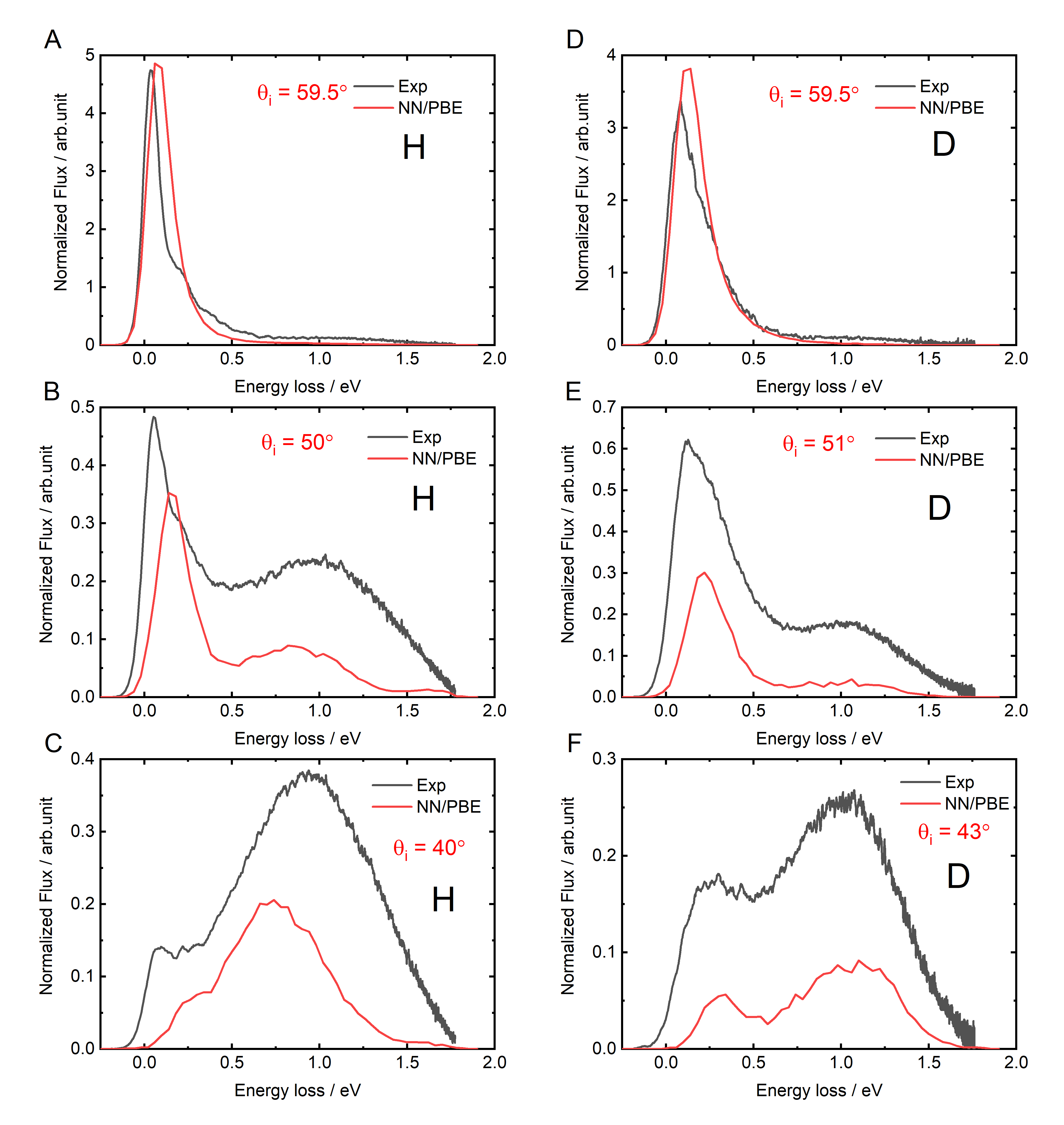}
    \caption{Comparing Theory with experiment: Angle integrated energy loss spectra. All incidence conditions are the same as in Fig.~\ref{Fig:Hangendist} }
    \label{Fig:elsHD}
\end{figure}

The quality of the results can be more clearly seen in angle-integrated energy loss distributions shown in Fig.~\ref{Fig:elsHD}. Here, the H and D energy loss distributions have in each case been normalized to the integrated scattering intensity of the $\theta_i=59.5^{\circ}$ distributions. The integrated scattering intensity drops off somewhat too rapidly with decreasing incidence angle, again likely reflecting the influence of out-of-plane scattering. The theoretically predicted energy loss in the quasi-elastic channel is somewhat larger than seen in experiment and the error is larger for H than for D. This might be a quantum effect allowing the H atom to sample the PES closer to the chemisorption barrier producing more inelasticity. 

The theoretically predicted D-atom energy loss for the transient bond-forming channel matches experiment remarkably well{---}see in particular panel D{---}but the H energy loss is smaller than experimental observation. We note that this is consistent with possible electronically non-adiabatic dynamics where the H atom energy loss is larger than that of the D-atom due to its higher velocity. 


\section{Conclusions}\label{sec:conclusions}					
We have developed a high-dimensional neural network potential energy surface for H- and D-atoms interacting with a free standing graphene sheet. The potential reproduces a large set of DFT-GGA electronic structure data with high accuracy and is sufficiently efficient to be used in large-scale molecular dynamics simulations. By computing several hundred thousand classical trajectories we demonstrated the utility of the PES by simulating angle- and energy-resolved H- and D-atom scattering experiments similar to those recently published\cite{Jiang2019}. 

The theoretical distributions are remarkably close to those seen in experiment. They accurately capture the branching between a quasi-elastic channel that samples only the physisorption well and a high-energy-loss channel that results from trajectories that traverse the chemisorption well.  The simulations also capture subtle differences between H and D scattering seen in experiment that appear as broadening of the angular distribution at specific values of $\theta_i$.  These results suggest that for scattering at 1.92\,eV, neglecting the Pt substrate in the model of scattering dynamics does not introduce large errors. 

We do, however, still see systematic differences between experiment and theory. These may be due to failure of the classical or Born--Oppenheimer approximations or both. It is, of course, possible that improved electronic structure data as well as a proper inclusion of the influence of the Pt substrate could explain the remaining discrepancies between experiment and theory. While the present work is only a first step, it demonstrates a crucial milestone toward developing a first principles quality PES that includes the influence of the metal substrate.

\section*{Conflicts of interest}
There are no conflicts to declare.

\section*{Acknowledgements}

We thank the DFG for financial support via the SFB 1073 projects A04 and C03 (Project No.\ 217133147). We gratefully acknowledge the funding of this project by computing time provided by the DFG project INST186/1294-1 FUGG (Project No.\ 405832858).



\balance


\bibliography{rsc} 

\providecommand*{\mcitethebibliography}{\thebibliography}
\csname @ifundefined\endcsname{endmcitethebibliography}
{\let\endmcitethebibliography\endthebibliography}{}
\begin{mcitethebibliography}{45}
\providecommand*{\natexlab}[1]{#1}
\providecommand*{\mciteSetBstSublistMode}[1]{}
\providecommand*{\mciteSetBstMaxWidthForm}[2]{}
\providecommand*{\mciteBstWouldAddEndPuncttrue}
  {\def\EndOfBibitem{\unskip.}}
\providecommand*{\mciteBstWouldAddEndPunctfalse}
  {\let\EndOfBibitem\relax}
\providecommand*{\mciteSetBstMidEndSepPunct}[3]{}
\providecommand*{\mciteSetBstSublistLabelBeginEnd}[3]{}
\providecommand*{\EndOfBibitem}{}
\mciteSetBstSublistMode{f}
\mciteSetBstMaxWidthForm{subitem}
{(\emph{\alph{mcitesubitemcount}})}
\mciteSetBstSublistLabelBeginEnd{\mcitemaxwidthsubitemform\space}
{\relax}{\relax}

\bibitem[Schlapbach and Zuttel({2001})]{Schlapbach2001}
L.~Schlapbach and A.~Zuttel, \emph{{NATURE}}, {2001}, \textbf{{414}},
  {353--358}\relax
\mciteBstWouldAddEndPuncttrue
\mciteSetBstMidEndSepPunct{\mcitedefaultmidpunct}
{\mcitedefaultendpunct}{\mcitedefaultseppunct}\relax
\EndOfBibitem
\bibitem[Hornekaer \emph{et~al.}({2003})Hornekaer, Baurichter, Petrunin, Field,
  and Luntz]{Hornekaer2003}
L.~Hornekaer, A.~Baurichter, V.~Petrunin, D.~Field and A.~Luntz,
  \emph{{SCIENCE}}, {2003}, \textbf{{302}}, {1943--1946}\relax
\mciteBstWouldAddEndPuncttrue
\mciteSetBstMidEndSepPunct{\mcitedefaultmidpunct}
{\mcitedefaultendpunct}{\mcitedefaultseppunct}\relax
\EndOfBibitem
\bibitem[Balog \emph{et~al.}({2010})Balog, Jorgensen, Nilsson, Andersen,
  Rienks, Bianchi, Fanetti, Laegsgaard, Baraldi, Lizzit, Sljivancanin,
  Besenbacher, Hammer, Pedersen, Hofmann, and Hornekaer]{Balog2010}
R.~Balog, B.~Jorgensen, L.~Nilsson, M.~Andersen, E.~Rienks, M.~Bianchi,
  M.~Fanetti, E.~Laegsgaard, A.~Baraldi, S.~Lizzit, Z.~Sljivancanin,
  F.~Besenbacher, B.~Hammer, T.~G. Pedersen, P.~Hofmann and L.~Hornekaer,
  \emph{{NATURE MATERIALS}}, {2010}, \textbf{{9}}, {315--319}\relax
\mciteBstWouldAddEndPuncttrue
\mciteSetBstMidEndSepPunct{\mcitedefaultmidpunct}
{\mcitedefaultendpunct}{\mcitedefaultseppunct}\relax
\EndOfBibitem
\bibitem[Fornace \emph{et~al.}({2015})Fornace, Lee, Miyamoto, Manby, and
  Miller]{Fornace2015}
M.~E. Fornace, J.~Lee, K.~Miyamoto, F.~R. Manby and T.~F. Miller, III,
  \emph{{JOURNAL OF CHEMICAL THEORY AND COMPUTATION}}, {2015}, \textbf{{11}},
  {568--580}\relax
\mciteBstWouldAddEndPuncttrue
\mciteSetBstMidEndSepPunct{\mcitedefaultmidpunct}
{\mcitedefaultendpunct}{\mcitedefaultseppunct}\relax
\EndOfBibitem
\bibitem[Ding \emph{et~al.}({2017})Ding, Tsuchiya, Manby, and
  Miller]{Ding2017a}
F.~Ding, T.~Tsuchiya, F.~R. Manby and T.~F. Miller, III, \emph{{JOURNAL OF
  CHEMICAL THEORY AND COMPUTATION}}, {2017}, \textbf{{13}}, {4216--4227}\relax
\mciteBstWouldAddEndPuncttrue
\mciteSetBstMidEndSepPunct{\mcitedefaultmidpunct}
{\mcitedefaultendpunct}{\mcitedefaultseppunct}\relax
\EndOfBibitem
\bibitem[Ding \emph{et~al.}({2017})Ding, Manby, and Miller]{Ding2017b}
F.~Ding, F.~R. Manby and T.~F. Miller, III, \emph{{JOURNAL OF CHEMICAL THEORY
  AND COMPUTATION}}, {2017}, \textbf{{13}}, {1605--1615}\relax
\mciteBstWouldAddEndPuncttrue
\mciteSetBstMidEndSepPunct{\mcitedefaultmidpunct}
{\mcitedefaultendpunct}{\mcitedefaultseppunct}\relax
\EndOfBibitem
\bibitem[Brenner \emph{et~al.}({2002})Brenner, Shenderova, Harrison, Stuart,
  Ni, and Sinnott]{Brenner2002}
D.~Brenner, O.~Shenderova, J.~Harrison, S.~Stuart, B.~Ni and S.~Sinnott,
  \emph{{JOURNAL OF PHYSICS-CONDENSED MATTER}}, {2002}, \textbf{{14}},
  {783--802}\relax
\mciteBstWouldAddEndPuncttrue
\mciteSetBstMidEndSepPunct{\mcitedefaultmidpunct}
{\mcitedefaultendpunct}{\mcitedefaultseppunct}\relax
\EndOfBibitem
\bibitem[Jiang \emph{et~al.}({2019})Jiang, Kammler, Ding, Dorenkamp, Manby,
  Wodtke, Miller, Kandratsenka, and B\"unermann]{Jiang2019}
H.~Jiang, M.~Kammler, F.~Ding, Y.~Dorenkamp, F.~R. Manby, A.~M. Wodtke, T.~F.
  Miller, III, A.~Kandratsenka and O.~B\"unermann, \emph{{SCIENCE}}, {2019},
  \textbf{{364}}, {379--382}\relax
\mciteBstWouldAddEndPuncttrue
\mciteSetBstMidEndSepPunct{\mcitedefaultmidpunct}
{\mcitedefaultendpunct}{\mcitedefaultseppunct}\relax
\EndOfBibitem
\bibitem[Behler(2016)]{P4885}
J.~Behler, \emph{J. Chem. Phys.}, 2016, \textbf{145}, 170901\relax
\mciteBstWouldAddEndPuncttrue
\mciteSetBstMidEndSepPunct{\mcitedefaultmidpunct}
{\mcitedefaultendpunct}{\mcitedefaultseppunct}\relax
\EndOfBibitem
\bibitem[Noé \emph{et~al.}(2020)Noé, Tkatchenko, M\"uller, and
  Clementi]{P5793}
F.~Noé, A.~Tkatchenko, K.-R. M\"uller and C.~Clementi, \emph{Ann. Rev. Phys.
  Chem.}, 2020, \textbf{71}, 361--390\relax
\mciteBstWouldAddEndPuncttrue
\mciteSetBstMidEndSepPunct{\mcitedefaultmidpunct}
{\mcitedefaultendpunct}{\mcitedefaultseppunct}\relax
\EndOfBibitem
\bibitem[Rowe \emph{et~al.}(2018)Rowe, Cs\'anyi, Alf\`e, and
  Michaelides]{P5187}
P.~Rowe, G.~Cs\'anyi, D.~Alf\`e and A.~Michaelides, \emph{Phys. Rev. B}, 2018,
  \textbf{97}, 054303\relax
\mciteBstWouldAddEndPuncttrue
\mciteSetBstMidEndSepPunct{\mcitedefaultmidpunct}
{\mcitedefaultendpunct}{\mcitedefaultseppunct}\relax
\EndOfBibitem
\bibitem[Wen and Tadmor(2019)]{P5850}
M.~Wen and E.~B. Tadmor, \emph{Phys. Rev. B}, 2019, \textbf{100}, 195419\relax
\mciteBstWouldAddEndPuncttrue
\mciteSetBstMidEndSepPunct{\mcitedefaultmidpunct}
{\mcitedefaultendpunct}{\mcitedefaultseppunct}\relax
\EndOfBibitem
\bibitem[Khaliullin \emph{et~al.}(2010)Khaliullin, Eshet, K\"uhne, Behler, and
  Parrinello]{P2594}
R.~Z. Khaliullin, H.~Eshet, T.~D. K\"uhne, J.~Behler and M.~Parrinello,
  \emph{Phys. Rev. B}, 2010, \textbf{81}, 100103\relax
\mciteBstWouldAddEndPuncttrue
\mciteSetBstMidEndSepPunct{\mcitedefaultmidpunct}
{\mcitedefaultendpunct}{\mcitedefaultseppunct}\relax
\EndOfBibitem
\bibitem[Khaliullin \emph{et~al.}(2011)Khaliullin, Eshet, K\"uhne, Behler, and
  Parrinello]{P3007}
R.~Z. Khaliullin, H.~Eshet, T.~D. K\"uhne, J.~Behler and M.~Parrinello,
  \emph{Nature Materials}, 2011, \textbf{10}, 693--697\relax
\mciteBstWouldAddEndPuncttrue
\mciteSetBstMidEndSepPunct{\mcitedefaultmidpunct}
{\mcitedefaultendpunct}{\mcitedefaultseppunct}\relax
\EndOfBibitem
\bibitem[Deringer and Csanyi(2017)]{P4958}
V.~L. Deringer and G.~Csanyi, \emph{Phys. Rev. B}, 2017, \textbf{95},
  094203\relax
\mciteBstWouldAddEndPuncttrue
\mciteSetBstMidEndSepPunct{\mcitedefaultmidpunct}
{\mcitedefaultendpunct}{\mcitedefaultseppunct}\relax
\EndOfBibitem
\bibitem[Behler and Parrinello(2007)]{PRL2007}
J.~Behler and M.~Parrinello, \emph{Phys. Rev. Lett.}, 2007, \textbf{98},
  146401\relax
\mciteBstWouldAddEndPuncttrue
\mciteSetBstMidEndSepPunct{\mcitedefaultmidpunct}
{\mcitedefaultendpunct}{\mcitedefaultseppunct}\relax
\EndOfBibitem
\bibitem[Jiang \emph{et~al.}({2020})Jiang, Tao, Kammler, Ding, Wodtke,
  Kandratsenka, Miller~III, and B\"unermann]{jiang2020nuclear}
H.~Jiang, X.~Tao, M.~Kammler, F.~Ding, A.~M. Wodtke, A.~Kandratsenka, T.~F.
  Miller~III and O.~B\"unermann, \emph{{arXiv:2007.03372 [physics.chem-ph]}},
  {2020}\relax
\mciteBstWouldAddEndPuncttrue
\mciteSetBstMidEndSepPunct{\mcitedefaultmidpunct}
{\mcitedefaultendpunct}{\mcitedefaultseppunct}\relax
\EndOfBibitem
\bibitem[B\"unermann \emph{et~al.}({2018})B\"unermann, Jiang, Dorenkamp,
  Auerbach, and Wodtke]{Buenermann2018}
O.~B\"unermann, H.~Jiang, Y.~Dorenkamp, D.~J. Auerbach and A.~M. Wodtke,
  \emph{{Rev. Sci. Inst.}}, {2018}, \textbf{{89}}, {094101}\relax
\mciteBstWouldAddEndPuncttrue
\mciteSetBstMidEndSepPunct{\mcitedefaultmidpunct}
{\mcitedefaultendpunct}{\mcitedefaultseppunct}\relax
\EndOfBibitem
\bibitem[Behler(2011)]{P2882}
J.~Behler, \emph{J. Chem. Phys.}, 2011, \textbf{134}, 074106\relax
\mciteBstWouldAddEndPuncttrue
\mciteSetBstMidEndSepPunct{\mcitedefaultmidpunct}
{\mcitedefaultendpunct}{\mcitedefaultseppunct}\relax
\EndOfBibitem
\bibitem[Artrith and Behler(2012)]{P3114}
N.~Artrith and J.~Behler, \emph{Phys. Rev. B}, 2012, \textbf{85}, 045439\relax
\mciteBstWouldAddEndPuncttrue
\mciteSetBstMidEndSepPunct{\mcitedefaultmidpunct}
{\mcitedefaultendpunct}{\mcitedefaultseppunct}\relax
\EndOfBibitem
\bibitem[Behler(2015)]{P4444}
J.~Behler, \emph{Int. J. Quantum Chem.}, 2015, \textbf{115}, 1032--1050\relax
\mciteBstWouldAddEndPuncttrue
\mciteSetBstMidEndSepPunct{\mcitedefaultmidpunct}
{\mcitedefaultendpunct}{\mcitedefaultseppunct}\relax
\EndOfBibitem
\bibitem[Behler(2017)]{P5128}
J.~Behler, \emph{Angew. Chem. Int. Ed.}, 2017, \textbf{56}, 12828\relax
\mciteBstWouldAddEndPuncttrue
\mciteSetBstMidEndSepPunct{\mcitedefaultmidpunct}
{\mcitedefaultendpunct}{\mcitedefaultseppunct}\relax
\EndOfBibitem
\bibitem[Behler(2014)]{P4106}
J.~Behler, \emph{J. Phys.: Condens. Matter}, 2014, \textbf{26}, 183001\relax
\mciteBstWouldAddEndPuncttrue
\mciteSetBstMidEndSepPunct{\mcitedefaultmidpunct}
{\mcitedefaultendpunct}{\mcitedefaultseppunct}\relax
\EndOfBibitem
\bibitem[Kresse and Furthm\"uller(1996)]{P0156}
G.~Kresse and J.~Furthm\"uller, \emph{Phys. Rev. B}, 1996, \textbf{54},
  11169\relax
\mciteBstWouldAddEndPuncttrue
\mciteSetBstMidEndSepPunct{\mcitedefaultmidpunct}
{\mcitedefaultendpunct}{\mcitedefaultseppunct}\relax
\EndOfBibitem
\bibitem[Kresse(1995)]{Kresse1995}
G.~Kresse, \emph{J. Non-Cryst. Solids}, 1995, \textbf{192–193},
  222–229\relax
\mciteBstWouldAddEndPuncttrue
\mciteSetBstMidEndSepPunct{\mcitedefaultmidpunct}
{\mcitedefaultendpunct}{\mcitedefaultseppunct}\relax
\EndOfBibitem
\bibitem[Kresse and Furthm\"uller(1996)]{Kresse1996}
G.~Kresse and J.~Furthm\"uller, \emph{Comput. Mater. Sci.}, 1996, \textbf{6},
  15–50\relax
\mciteBstWouldAddEndPuncttrue
\mciteSetBstMidEndSepPunct{\mcitedefaultmidpunct}
{\mcitedefaultendpunct}{\mcitedefaultseppunct}\relax
\EndOfBibitem
\bibitem[Kresse and Joubert(1999)]{Kresse1999}
G.~Kresse and D.~Joubert, \emph{Phys. Rev. B}, 1999, \textbf{59},
  1758--1775\relax
\mciteBstWouldAddEndPuncttrue
\mciteSetBstMidEndSepPunct{\mcitedefaultmidpunct}
{\mcitedefaultendpunct}{\mcitedefaultseppunct}\relax
\EndOfBibitem
\bibitem[Perdew \emph{et~al.}(1996)Perdew, Burke, and Ernzerhof]{Perdew1996a}
J.~P. Perdew, K.~Burke and M.~Ernzerhof, \emph{Phys. Rev. Lett.}, 1996,
  \textbf{77}, 3865--3868\relax
\mciteBstWouldAddEndPuncttrue
\mciteSetBstMidEndSepPunct{\mcitedefaultmidpunct}
{\mcitedefaultendpunct}{\mcitedefaultseppunct}\relax
\EndOfBibitem
\bibitem[Grimme(2006)]{Grimme2006SemiempiricalGD}
S.~Grimme, \emph{Journal of computational chemistry}, 2006, \textbf{27 15},
  1787--99\relax
\mciteBstWouldAddEndPuncttrue
\mciteSetBstMidEndSepPunct{\mcitedefaultmidpunct}
{\mcitedefaultendpunct}{\mcitedefaultseppunct}\relax
\EndOfBibitem
\bibitem[Bl\"ochl(1994)]{PhysRevB.50.17953}
P.~E. Bl\"ochl, \emph{Phys. Rev. B}, 1994, \textbf{50}, 17953--17979\relax
\mciteBstWouldAddEndPuncttrue
\mciteSetBstMidEndSepPunct{\mcitedefaultmidpunct}
{\mcitedefaultendpunct}{\mcitedefaultseppunct}\relax
\EndOfBibitem
\bibitem[Monkhorst and Pack(1976)]{Monkhorst1976}
H.~J. Monkhorst and J.~D. Pack, \emph{Phys. Rev. B}, 1976, \textbf{13},
  5188--5192\relax
\mciteBstWouldAddEndPuncttrue
\mciteSetBstMidEndSepPunct{\mcitedefaultmidpunct}
{\mcitedefaultendpunct}{\mcitedefaultseppunct}\relax
\EndOfBibitem
\bibitem[Bl\"ochl \emph{et~al.}(1994)Bl\"ochl, Jepsen, and
  Andersen]{PhysRevB.49.16223}
P.~E. Bl\"ochl, O.~Jepsen and O.~K. Andersen, \emph{Phys. Rev. B}, 1994,
  \textbf{49}, 16223--16233\relax
\mciteBstWouldAddEndPuncttrue
\mciteSetBstMidEndSepPunct{\mcitedefaultmidpunct}
{\mcitedefaultendpunct}{\mcitedefaultseppunct}\relax
\EndOfBibitem
\bibitem[Stukowski({2010})]{ovito}
A.~Stukowski, \emph{{MODELLING AND SIMULATION IN MATERIALS SCIENCE AND
  ENGINEERING}}, {2010}, \textbf{{18}}, {015012}\relax
\mciteBstWouldAddEndPuncttrue
\mciteSetBstMidEndSepPunct{\mcitedefaultmidpunct}
{\mcitedefaultendpunct}{\mcitedefaultseppunct}\relax
\EndOfBibitem
\bibitem[Artrith and Behler(2012)]{Behler2012Cu}
N.~Artrith and J.~Behler, \emph{Phys. Rev. B}, 2012, \textbf{85}, 045439\relax
\mciteBstWouldAddEndPuncttrue
\mciteSetBstMidEndSepPunct{\mcitedefaultmidpunct}
{\mcitedefaultendpunct}{\mcitedefaultseppunct}\relax
\EndOfBibitem
\bibitem[Behler(Universit\"at G\"ottingen 2020)]{IPClink}
J.~Behler, \emph{Ru{NN}er - A Neural Network Code for High-Dimensional
  Potential-Energy Surfaces}, Universit\"at G\"ottingen 2020,
  \url{http://www.uni-goettingen.de/de/560580.html}\relax
\mciteBstWouldAddEndPuncttrue
\mciteSetBstMidEndSepPunct{\mcitedefaultmidpunct}
{\mcitedefaultendpunct}{\mcitedefaultseppunct}\relax
\EndOfBibitem
\bibitem[Blank and Brown(1994)]{P1308}
T.~B. Blank and S.~D. Brown, \emph{J. Chemometrics}, 1994, \textbf{8},
  391--407\relax
\mciteBstWouldAddEndPuncttrue
\mciteSetBstMidEndSepPunct{\mcitedefaultmidpunct}
{\mcitedefaultendpunct}{\mcitedefaultseppunct}\relax
\EndOfBibitem
\bibitem[Auerbach \emph{et~al.}(2020)Auerbach, Janke, Kammler, Kandratsenka,
  and Wille]{MDT2GIT}
D.~J. Auerbach, S.~M. Janke, M.~Kammler, S.~Kandratsenka and S.~Wille,
  \emph{Molecular Dynamics Tian Xia 2 (MDT2); program for simulating the
  scattering of atoms and molecules from a surface (GitHub repository).
  Available at https://github.com/akandra/md\_tian2}, 2020\relax
\mciteBstWouldAddEndPuncttrue
\mciteSetBstMidEndSepPunct{\mcitedefaultmidpunct}
{\mcitedefaultendpunct}{\mcitedefaultseppunct}\relax
\EndOfBibitem
\bibitem[Janke \emph{et~al.}({2015})Janke, Auerbach, Wodtke, and
  Kandratsenka]{Janke2015}
S.~M. Janke, D.~J. Auerbach, A.~M. Wodtke and A.~Kandratsenka, \emph{{JOURNAL
  OF CHEMICAL PHYSICS}}, {2015}, \textbf{{143}}, {124708}\relax
\mciteBstWouldAddEndPuncttrue
\mciteSetBstMidEndSepPunct{\mcitedefaultmidpunct}
{\mcitedefaultendpunct}{\mcitedefaultseppunct}\relax
\EndOfBibitem
\bibitem[B\"unermann \emph{et~al.}({2015})B\"unermann, Jiang, Dorenkamp,
  Kandratsenka, Janke, Auerbach, and Wodtke]{Buenermann2015}
O.~B\"unermann, H.~Jiang, Y.~Dorenkamp, A.~Kandratsenka, S.~M. Janke, D.~J.
  Auerbach and A.~M. Wodtke, \emph{{SCIENCE}}, {2015}, \textbf{{350}},
  {1346--1349}\relax
\mciteBstWouldAddEndPuncttrue
\mciteSetBstMidEndSepPunct{\mcitedefaultmidpunct}
{\mcitedefaultendpunct}{\mcitedefaultseppunct}\relax
\EndOfBibitem
\bibitem[Allen and Tildesley(1989)]{allen1989computer}
M.~P. Allen and D.~J. Tildesley, \emph{Computer Simulation of Liquids},
  Clarendon Press, USA, 1989\relax
\mciteBstWouldAddEndPuncttrue
\mciteSetBstMidEndSepPunct{\mcitedefaultmidpunct}
{\mcitedefaultendpunct}{\mcitedefaultseppunct}\relax
\EndOfBibitem
\bibitem[Swope \emph{et~al.}(1982)Swope, Andersen, Berens, and
  Wilson]{Swope1982}
W.~C. Swope, H.~C. Andersen, P.~H. Berens and K.~R. Wilson, \emph{J. Chem.
  Phys.}, 1982, \textbf{76}, 637\relax
\mciteBstWouldAddEndPuncttrue
\mciteSetBstMidEndSepPunct{\mcitedefaultmidpunct}
{\mcitedefaultendpunct}{\mcitedefaultseppunct}\relax
\EndOfBibitem
\bibitem[Frenkel and Smit(2002)]{dfrenkel96:mc}
D.~Frenkel and B.~Smit, \emph{Understanding Molecular Simulation: From
  Algorithms to Applications}, Academic Press, San Diego, 2nd edn, 2002,
  vol.~1\relax
\mciteBstWouldAddEndPuncttrue
\mciteSetBstMidEndSepPunct{\mcitedefaultmidpunct}
{\mcitedefaultendpunct}{\mcitedefaultseppunct}\relax
\EndOfBibitem
\bibitem[Andersen(1980)]{Andersen1980}
H.~C. Andersen, \emph{J. Chem. Phys.}, 1980, \textbf{72}, 2384\relax
\mciteBstWouldAddEndPuncttrue
\mciteSetBstMidEndSepPunct{\mcitedefaultmidpunct}
{\mcitedefaultendpunct}{\mcitedefaultseppunct}\relax
\EndOfBibitem
\bibitem[Ghio \emph{et~al.}(1980)Ghio, Mattera, Salvo, Tommasini, and
  Valbusa]{ghio1980vibrational}
E.~Ghio, L.~Mattera, C.~Salvo, F.~Tommasini and U.~Valbusa, \emph{The Journal
  of chemical physics}, 1980, \textbf{73}, 556--561\relax
\mciteBstWouldAddEndPuncttrue
\mciteSetBstMidEndSepPunct{\mcitedefaultmidpunct}
{\mcitedefaultendpunct}{\mcitedefaultseppunct}\relax
\EndOfBibitem
\bibitem[Bonfanti \emph{et~al.}(2007)Bonfanti, Martinazzo, Tantardini, and
  Ponti]{Bonfanti2007}
M.~Bonfanti, R.~Martinazzo, G.~F. Tantardini and A.~Ponti, \emph{The Journal of
  Physical Chemistry C}, 2007, \textbf{111}, 5825--5829\relax
\mciteBstWouldAddEndPuncttrue
\mciteSetBstMidEndSepPunct{\mcitedefaultmidpunct}
{\mcitedefaultendpunct}{\mcitedefaultseppunct}\relax
\EndOfBibitem
\end{mcitethebibliography}
\bibliographystyle{rsc} 

\end{document}


\pagestyle{fancy}
\thispagestyle{plain}
\fancypagestyle{plain}{
\renewcommand{\headrulewidth}{0pt}
}

\makeFNbottom
\makeatletter
\renewcommand\LARGE{\@setfontsize\LARGE{15pt}{17}}
\renewcommand\Large{\@setfontsize\Large{12pt}{14}}
\renewcommand\large{\@setfontsize\large{10pt}{12}}
\renewcommand\footnotesize{\@setfontsize\footnotesize{7pt}{10}}
\makeatother

\renewcommand{\thefootnote}{\fnsymbol{footnote}}
\renewcommand\footnoterule{\vspace*{1pt}%
\color{cream}\hrule width 3.5in height 0.4pt \color{black}\vspace*{5pt}} 
\setcounter{secnumdepth}{5}

\makeatletter 
\renewcommand\@biblabel[1]{#1}            
\renewcommand\@makefntext[1]%
{\noindent\makebox[0pt][r]{\@thefnmark\,}#1}
\makeatother 
\renewcommand{\figurename}{\small{Fig.}~}
\sectionfont{\sffamily\Large}
\subsectionfont{\normalsize}
\subsubsectionfont{\bf}
\setstretch{1.125} 
\setlength{\skip\footins}{0.8cm}
\setlength{\footnotesep}{0.25cm}
\setlength{\jot}{10pt}
\titlespacing*{\section}{0pt}{4pt}{4pt}
\titlespacing*{\subsection}{0pt}{15pt}{1pt}

\renewcommand{\headrulewidth}{0pt} 
\renewcommand{\footrulewidth}{0pt}
\setlength{\arrayrulewidth}{1pt}
\setlength{\columnsep}{6.5mm}
\setlength\bibsep{1pt}

\makeatletter 
\newlength{\figrulesep} 
\setlength{\figrulesep}{0.5\textfloatsep} 

\newcommand{\topfigrule}{\vspace*{-1pt}%
\noindent{\color{cream}\rule[-\figrulesep]{\columnwidth}{1.5pt}} }

\newcommand{\botfigrule}{\vspace*{-2pt}%
\noindent{\color{cream}\rule[\figrulesep]{\columnwidth}{1.5pt}} }

\newcommand{\dblfigrule}{\vspace*{-1pt}%
\noindent{\color{cream}\rule[-\figrulesep]{\textwidth}{1.5pt}} }

\makeatother

\twocolumn[
  \begin{@twocolumnfalse}
\begin{tabular}{m{4.5cm} p{13.5cm} }

& \noindent\LARGE{\textbf{An experimentally validated neural-network potential energy surface for H atoms on free-standing graphene in full dimensionality$^\dag$}} \\
\vspace{0.3cm} & \vspace{0.3cm} \\

 & \noindent\large{Sebastian Wille\textit{$^{a,c}$}, Hongyan Jiang\textit{$^{a}$}, Oliver B\"unermann\textit{$^{a,b,d}$}, Alec~M.~Wodtke\textit{$^{a,b,d}$}, J\"org Behler\textit{$^{c,d}$}, and Alexander Kandratsenka$^{\ast}$\textit{$^{a}$}} \\

\end{tabular}

 \end{@twocolumnfalse} \vspace{0.6cm}

  ]

\renewcommand*\rmdefault{bch}\normalfont\upshape
\rmfamily
\section*{}
\vspace{-1cm}


\footnotetext{\textit{$^{a}$~Department of Dynamics at Surfaces, Max-Planck-Institute for Biophysical Chemistry, Am Fa{\ss}berg 11, 37077 G{\"o}ttingen, Germany. E-mail: akandra@mpibpc.mpg.de}}
\footnotetext{\textit{$^{b}$~Institute for Physical Chemistry, Georg-August University of G\"ottingen, Tammann-stra\ss e 6, 37077 G\"ottingen, Germany}}
\footnotetext{\textit{$^{c}$~Institute for Physical Chemistry, Theoretische Chemie, Georg-August University of G\"ottingen, Tammannstra\ss e 6, 37077 G\"ottingen, Germany.}}
\footnotetext{\textit{$^{d}$~International Center for Advanced Studies of Energy Conversion, Georg-August University of G\"ottingen, Tammannstra\ss e 6, 37077 G\"ottingen, Germany.}}

\footnotetext{\dag~Electronic Supplementary Information (ESI) available: [details of any supplementary information available should be included here]. See DOI: 10.1039/cXCP00000x/}



\section*{SUPPLEMENTARY INFORMATION}\label{sec:SI}

\section{HDNNP details}\label{sec:SInn}

The NNs for hydrogen and carbon atoms were constructed using symmetry functions of radial type
\[
\rho_m = \sum_n f_\text{c}(r_{mn}) \exp\left[-\eta r_{mn}^2\right],
\]
and angular type
\[
\phi_m =\sum_{k,n\neq m}  \frac{(1+\lambda\cos\theta_{kmn})^\zeta}{2^{\zeta-1}} f_\text{c}(r_{km})f_\text{c}(r_{mn})f_\text{c}(r_{kn}) e^{-\eta (r_{km}^2 + r_{mn}^2 + r_{kn}^2)}
\]
centered on atom $m$. Here $r_{mn}$ denotes the distance between atoms $m$ and $n$, $\theta_{kmn}$ is the angle between vectors $\mathbf{r}_{mk}$ and $\mathbf{r}_{mn}$; $f_\text{c}$ is a cutoff function, which gives zero if its argument is larger than $12 a_0$ and 1 otherwise. Indices in sums run over all the neighboring atoms of central atom $m$. $\eta$, $\lambda$ and $\zeta$ are the parameters defining a symmetry function, their values are listed in Table~\ref{TAB:SymFunc}. 
We use 30 input neurons for an H-atom and 60 per C-atom.

\begin{table} [h!]
\caption{Parameters for the atom-centered symmetry functions for H- and C-atoms. Equations for the types used can be found in section~\ref{sec:SInn}.}
\begin{center}
\begin{tabular}{ r  r  r  r }
\toprule
 no. & $\eta (a_0^{-2})$ & $\lambda$ & $\zeta$ \\
\midrule
\multicolumn{4}{l}{Radial}\\
1 & 0.000  &  &  \\
2 & 0.005  &  &  \\
3 & 0.013  &  &  \\
4 & 0.027  &  &  \\
5 & 0.060  &  &  \\
6 & 0.156  &  &  \\
\multicolumn{4}{l}{Angular}\\
7 & 0.000 & 1 & 1   \\
8 & 0.000 & 1 & 2   \\
9 & 0.000 & 1 & 4   \\
10& 0.000 & 1 & 16   \\
11& 0.000 & -1 & 1   \\
12& 0.000 & -1 & 2   \\
13& 0.000 & -1 & 4   \\
14& 0.000 & -1 & 16   \\
15& 0.013 & 1 & 1   \\
16& 0.013 & 1 & 2   \\
17& 0.013 & 1 & 4   \\
18& 0.013 & 1 & 16   \\
19& 0.013 & -1 & 1   \\
20& 0.013 & -1 & 2   \\
21& 0.013 & -1 & 4   \\
22& 0.013 & -1 & 16   \\
15& 0.156 & 1 & 1   \\
16& 0.156 & 1 & 2   \\
17& 0.156 & 1 & 4   \\
18& 0.156 & 1 & 16   \\
19& 0.156 & -1 & 1   \\
20& 0.156 & -1 & 2   \\
21& 0.156 & -1 & 4   \\
22& 0.156 & -1 & 16   \\
\bottomrule
\end{tabular}
\end{center}
\label{TAB:SymFunc}
\end{table}

\begin{figure}[h!]
    \center
\includegraphics[width=1.0\columnwidth]{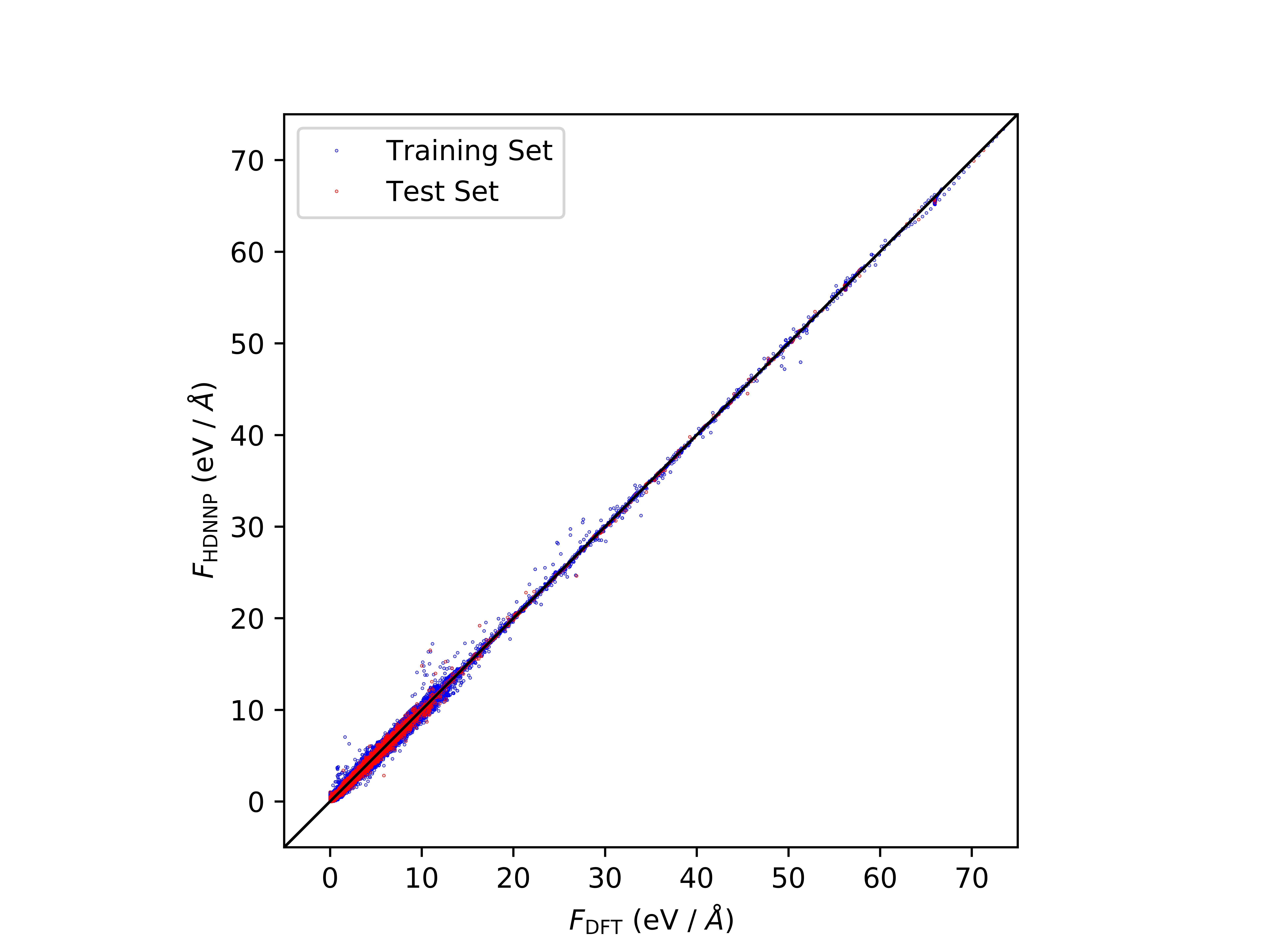}
    \caption{Correlation plot of HDNNP $|F_\text{HDNN-PES}|$ and DFT forces $|F_\text{DFT}|$ for the training set in blue and test set in red, respectively. The HDNNP represents the DFT forces over the whole range quite well or at least reasonable.}
    \label{fig:forcecorr}
\end{figure}

\begin{figure}[h!]
    \center
\includegraphics[width=1.0\columnwidth]{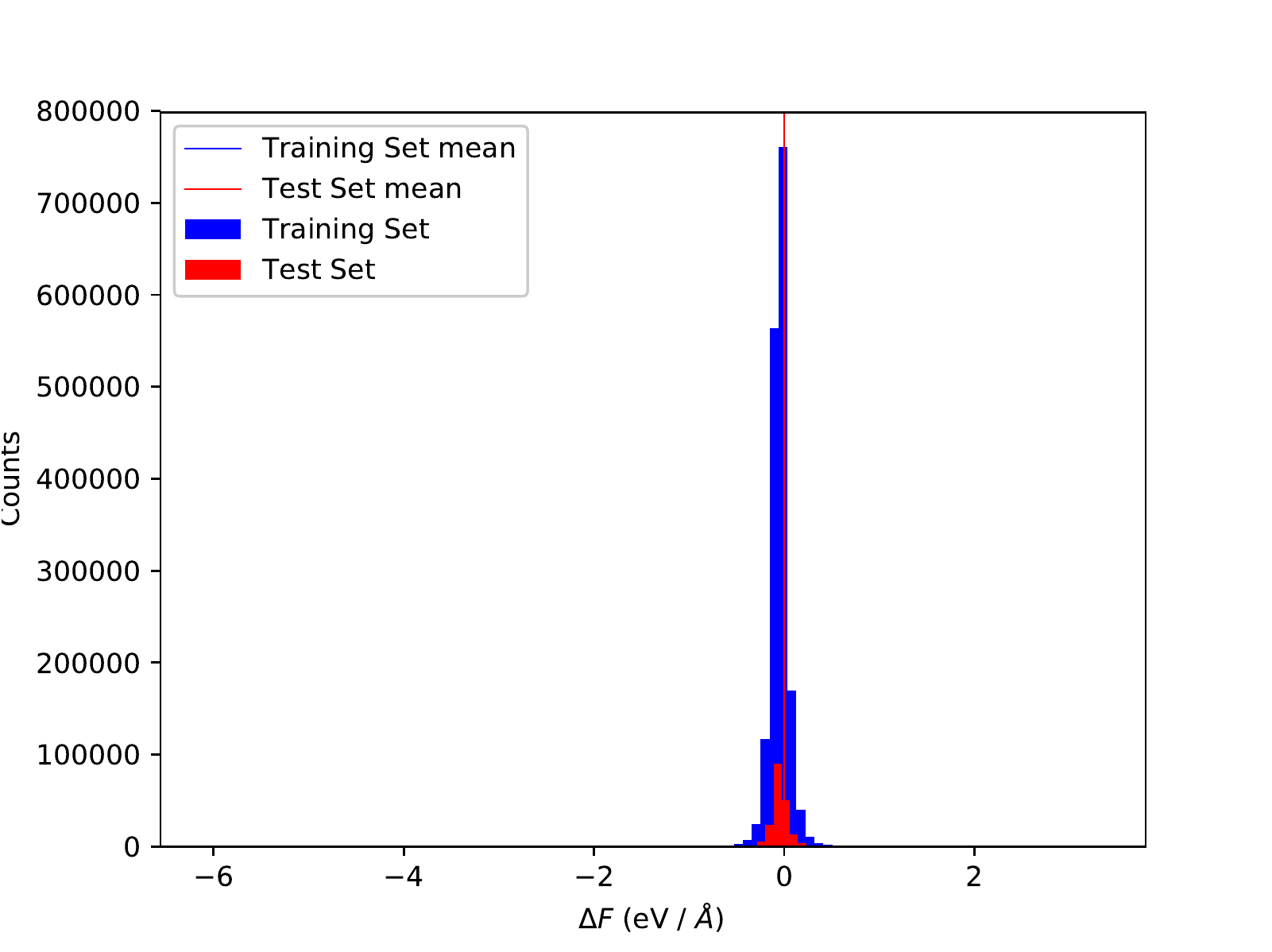}
    \caption{Histogram of differences of the forces $F_\text{DFT} - F_\text{HDNN-PES}$ for the training set in blue and test set in red, respectively. The mean values for the training and test set are also shown in the corresponding colour.}
    \label{fig:forcehist}
\end{figure}

The quality of the fit to the DFT data is shown for the energies in Fig.~3 of the main body of the paper.
In this supplementary section, Fig.~\ref{fig:forcecorr} shows in turn the correlation of the amplitudes of forces $|F_\text{DFT}|$ extracted from the DFT data and forces $|F_\text{NN}|$ obtained with HDNNP. Fig.~\ref{fig:forcehist} presents the same information in the form of the histogram providing a clearer representation of the fit errors. 
The RMSE associated with the forces is small, $\sim$90\,meV/\AA.






\section{MD Trajectories}

The most important performance feature of the HDNN-PES is its ability to accurately calculate the energy and forces for a system with many degrees of freedom with low computational costs. This is particularly important in the simulations of angle resolved atomic scattering experiments from surfaces. The experimental design detects only a small fraction of the scattered atoms due to its high angular resolution. Consequently, one needs hundreds of thousand trajectories to reduce statistical noise to the level of the experimental data. Table~\ref{tab:trajs} shows
the total number of MD trajectories calculated at the conditions used in the paper as well as the number of trajectories scattered into the cone with the apex angle of $3^\circ$ corresponding to the geometry of the experimental setup.

\begin{table} [b!]
\caption{Showing the total number of simulated trajectories $N_\text{total}$, scattered trajectories within detection limit compared to experimental setup $N_{3^\circ}$, the normal component of incidence energy and sticking probability $S_0$ for H/D at incidence polar angle $\theta_i$.}
\begin{center}
\begin{tabular}{ l  l  l  l  l  l  l }
\toprule
 & $\theta_i$ & $N_\text{total}$  & $N_{3^\circ}$ & $E_i\cos^2\theta_i$/eV & $S_0$ \\
\midrule
H &	40 & 354,911 & 22,630 & 1.13 & 0.21 \\
 & 50 & 291,238 & 22,533 & 0.79 & 0.39 \\
 & 59.5 & 322,691 & 122,205 & 0.49 & 0.22 \\
D & 43 & 189,577 & 7,021 & 1.03 & 0.52 \\
 & 51 & 183,837 & 10,452 & 0.76 & 0.68 \\
 & 59.5 & 221,861 & 73,607 & 0.49 & 0.39 \\
\bottomrule
\end{tabular}
\end{center}
\label{tab:trajs}
\end{table}

When simulating hydrogen scattering from graphene, the trajectories were interrupted after a 200\,fs simulation time. H atoms which had not left the surface after this time were considered to have adsorbed. The sticking probabilities for the incidence conditions used in the main paper are collected in Table~\ref{tab:trajs}.


\section{Trajectories visualized}

\begin{figure}[t!]
\includegraphics[width=1.0\columnwidth]{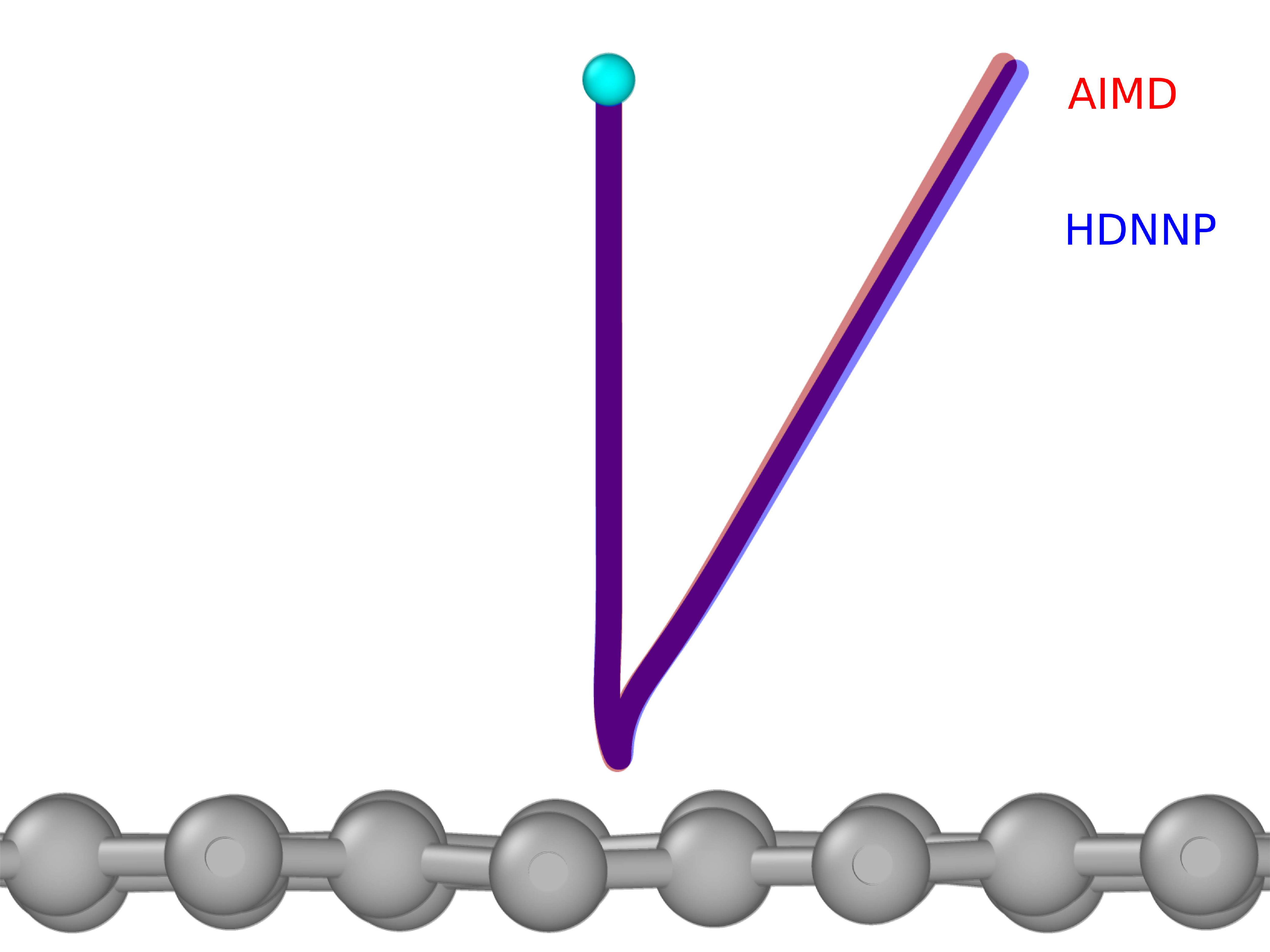}
    \caption{Side view of Fig.~5 of the main text.}
    \label{FIG:AIMDTRAJ_side}
\end{figure}

To get a feeling how the scattering happens it is often useful to visualize trajectory as shown in Fig.~5 on the main body of the paper. Fig.~\ref{FIG:AIMDTRAJ_side} provides a side view additional to the top view shown in Fig.~5 of the main text. We also created several movies, where one can follow the trajectories in more detail. We added a top and a side view for the AIMD and HDNNP trajectory. Furthermore, we added three movies showing an example of the quasi-elastic energy loss channel (fast channel), the high energy loss channel (slow channel) and adsorption of the projectile on the surface, which is not possible to see in the experiment.
All the movies included in the SI are created using OVITO version 2.9.0.

\balance
